\documentclass[usenatbib]{mn2e}
\usepackage{rotating}
\usepackage{booktabs}
\usepackage{setspace}
\usepackage{txfonts}
\usepackage{float}
\usepackage{amssymb}
\usepackage{epsfig}
\usepackage{color}
\usepackage{graphicx}
\usepackage{multirow}
\usepackage{longtable}
\usepackage{verbatim}
\usepackage[flushleft]{threeparttable}
\usepackage{gensymb}
\usepackage{epstopdf}
\usepackage{array}
\usepackage{natbib}

\newcommand{\Nu}{\it{NuSTAR}}
\newcommand{\sw}{\it{Swift}}
\newcommand{\IGR}{IGR~J17091}
\newcommand{\rgs}{GRS~1915+105}

\title[The reflection spectrum of IGR~J17091--3642]{The reflection component in the average and heartbeat spectra of the black-hole candidate IGR~J17091--3642 during the 2016 outburst}

\author[Yanan Wang et al.]{Yanan Wang$^{1}$\thanks{E-mail: yanan@astro.rug.nl}, Mariano M\'endez$^{1}$, Diego Altamirano$^{2}$, James Court$^{2}$, Aru Beri$^{2}$	 \newauthor
and Zheng Cheng$^{1}$\\
$^{1}$Kapteyn Astronomical Institute, University of Groningen, PO Box 800, 9700 AV Groningen, The Netherlands\\
$^{2}$Physics \& Astronomy, University of Southampton, Southampton, Hampshire SO17 1BJ, UK\\}

\begin{document}
\maketitle
%
%
\begin{abstract}
We present simultaneous {\Nu} and {\sw} observations of the black hole transient IGR~J17091--3642 during its 2016 outburst.
By jointly fitting six {\Nu} and four {\sw} spectra, we found that during this outburst the source evolves from the hard to the hard/soft intermediate and back to the hard state, similar to the 2011 outburst. 
Unlike in the previous outburst, in this case we observed both a broad emission and an moderately broad absorption line in our observations. Our fits favour an accretion disc with an inclination angle of $\sim$$45\degree$ with respect to the line of sight and a high iron abundance of $3.5\pm0.3$ in units of the solar abundance. 
We also observed heartbeat variability in one {\Nu} observation. We fitted the phase-resolved spectra of this observation and found that the reflected emission varies independently from the direct emission, whereas in the fits to the average spectra these two quantities are strongly correlated.
Assuming that in IGR~J17091--3642 the inner radius of the disc both in the average and the phase-resolved spectra is located at the radius of the innermost stable circular orbit, with 90\% confidence the spin parameter of the black hole in this system is $-0.13\leq a_{*}\leq0.27$. 

\end{abstract}


\section{Introduction}\label{sec:intro}
In low mass X-ray binaries (LMXBs), the main source of power is the gravitational energy released by matter accreted from a companion star onto a compact object, a black hole or a neutron star (e.g., \citealt*{Frank2002}).
Based on the long-term evolution of the X-ray emission, LMXBs can be classified into persistent and transient sources.
Apart from a few exceptions (e.g., Cyg~X--1 and {\rgs}), LMXBs containing a black hole (BH) primary are transients (see, e.g., \citealt{Tanaka1996}; \citealt*{Chen1997}; \citealt{Tomsick2000} for a comprehensive description of this class of sources). 
These transients are most of the time in a dim (emitting luminosities below $\rm 10^{33}~erg~s^{-1}$) and quiescent state but, occasionally, they display outbursts where $L_{x}$ can 
increase by several orders of magnitude in a few days, and the source remains bright for weeks to months. The recurrence time between outbursts varies from months to decades (see e.g. \citealt*{Coriat2012}; \citealt{Tetarenko2016}).

Using timing and spectral properties of black hole X-ray binaries (BHXBs), in the past three decades it was discovered that these systems recurrently follow a common spectral and timing pattern during outbursts,
possibly driven by transitions between different accretion regimes (e.g., \citealt{Miyamoto1991,Klis1995,Mariano1997}; \citealt*{Belloni2011}). In general, an outburst starts with the source in the hard state, associated with strong broadband (0.1$-$40~Hz) variability and an inverse Compton component that dominates the spectrum in the 1$-$20~keV range. As the outburst progresses, the source transits through the hard and the soft intermediate states, into the soft state, in which the spectrum softens, the thermal component dominates the emission, and the strength of the broadband variability decreases. 
At the end of the outburst, the spectrum hardens again and, at lower luminosities than in the rising phase of the outburst, the source moves back to the intermediate and the hard states, until the source becomes too dim to be detected \citep[e.g.,][]{Belloni2000,Homan2001}.

IGR~J17091--3642 (hereafter, {\IGR}) was discovered with {\it INTEGRAL}/IBIS in April 2003 \citep{Kuulkers2003}, and was subsequently detected during outburst in 2007, 2011 and 2016 \citep[e.g.,][]{Capitanio2009,Krimm2011,Miller2016}. 
Similar to GRS~1915+105, a so-called micro-quasar with unique X-ray `variability classes' \citep{Belloni2000}, {\IGR} displays a number of variability classes which are identified by quasi-periodic flares, dips and high-amplitude intensity variations \citep{Altamirano2011,Altamirano2012}. The most prominent and best-studied of these patterns is the highly regular flaring heartbeat class, similar to the $\rho$ class in GRS~1915+105 \citep{Belloni2000}. Despite the similarities, \cite{Court2017} suggested that the heartbeat variability in these two objects may be the result of different physical processes.

Spectral analysis provides a useful way to probe the physics and geometry of X-ray binaries.
In most cases, the spectral analysis is carried out on the average spectrum of an observation of a source. 
In sources with periodic or quasi-periodic variability that is shorter than the length of an observation, 
phase-resolved spectroscopic studies are more appropriate to reveal which spectral components cause the variability.

Although there are different methods to carry out phase-resolved spectroscopy (e.g. \citealt*{Belloni2001}; \citealt{Miller2005,Wilkinson2011,Ingram2015}), some of them are only applicable to either a bright source or a well-defined quasi-periodic oscillation (QPO) waveform.
\cite{Stevens2016} developed a sophisticated technique that involves cross-correlating each energy channel with a reference band without the requirements of a periodic signal or high count rate. 
However, due to the limitations on the data quality, the modulation in the iron line energy could not be measured in their work.
\cite{Court2017} developed a variable-period folding algorithm for phase-resolved analysis, for the case of high-amplitude QPOs that show significant frequency shifts on time-scales shorter than the length of an observation.

Very recently, \cite{Xu2017} presented a spectral and timing study of three {\Nu} and {\sw} observations of {\IGR} in the hard state during the 2016 outburst, and they found a disc reflection component in all the three {\Nu} spectra which they fitted with relativistic reflection models \citep{Dauser2014,Garc2014}. 
In this paper we analyse all six {\Nu} and four simultaneous {\sw} observations during the same outburst of {\IGR} to study the spectral evolution along this outburst and we also carried out a phase-resolved analysis of one {\Nu} observation with heartbeat variability. This paper is organized as follows: In Section~2 we describe the observations and analysis methods. Our results of the average and the phase-resolved spectra are presented in Section~3, followed by a discussion in Section~4, and we summarize our conclusions in Section~5.

\section{Observations and data analysis}\label{sec:dataanalysis}
On 26 Feb 2016, observations with {\sw}/BAT (Burst Alert Telescope, \citealt{Barthelmy2000}) indicated that {\IGR} was in outburst \citep{Miller2016}; {\sw}/XRT (X-ray Telescope, \citealt{Burrows2003}) was later on used to follow up the entire outburst, which lasted for almost 200~days, between MJDs 57444 and 57640. We created a long-term 0.3$-$10~keV {\sw}/XRT light curve of the 2016 outburst of {\IGR}, as shown in Fig.~1a, using the on-line light-curve generator provided by the UK Swift Science Data Centre (UKSSDC; \citealt{Evans2007}).

During this outburst, {\IGR} was observed six times with {\Nu}, between March 12 and May 26 2016. We refer to these six observations as NS1, NS2, NS3, N4, N5 and NS6 here (see details in Table~1); the red arrows in Fig.~1 mark the times of the {\Nu} observations discussed in this work.
To analyse the {\Nu} data, we created light curves and spectra of the source and the background with the command {\it nuproducts} from the {\Nu} Data Analysis Software (NuSTARDAS) version 1.9.1 \citep{Harrison2013}, using a circular extraction region of 100\arcsec~centered at the source, and another region of the same size centred away from the source for the background. 

For five of the six {\Nu} observations, {\IGR} was simultaneously observed with {\sw}/XRT in Windowed Timing mode (see Table~1). 
The {\sw} observations correspond to the green arrows in Fig.~1a.
We created the corresponding source and background light curves, spectra, and response matrices using the on-line spectrum generator provided by UKSSDC.
Finally we re-binned all the {\Nu} and {\sw} spectra using the task {\it grppha} within {\it ftools}\footnote{https://heasarc.gsfc.nasa.gov/ftools/} to reach a minimum of 25 counts per spectral bin.

The spectral fitting was done using XSPEC (12.9.1a).
We fitted the {\Nu} spectra from the FPMA and FPMB instruments simultaneously from 3 to 75~keV with different models, including a multiplicative term fixed to 1 for the FPMA and allowed to float for the FPMB. The value of this term varies within 0.99$-$1.02, indicating that the relative calibration of the two instruments is within $\pm$2\%. 
While in general, the X-ray spectra of two instruments should not be combined but fitted individually, the systematic errors in the calibration of the {\Nu}
telescope is smaller than the statistical errors in the data of {\IGR}, and all our fits of the data jointly or combined gave the same value of the parameters within errors. Therefore, for the rest of the paper, we combined the FPMA and FPMB spectra in each {\Nu} observation to speed up the fitting process.

We fitted the {\sw} data only in the energy range of 0.8$-$3~keV
to estimate the column density and the possible emission from a soft component in this work. 
To account for the interstellar absorption, in all fits we used the component {\sc phabs} with the abundance table of \citet*{Wilms2000} and the cross section table of \cite{Verner1996}. Unless explicitly mentioned, we quote all errors at 1-$\sigma$ confidence level.

\begin{table*} 
\caption{{\Nu} and {\sw} observations of {\IGR} used in this paper}
\renewcommand{\arraystretch}{1.3}
\footnotesize
\centering
\begin{tabular}{lccccc}
\hline \hline
\multirow{2}{*}{Observations}&\multirow{2}{*}{Mission}&\multirow{2}{*}{Identification Number}&Start Times &\multirow{2}{*}{Exposure (ks)}\\
                        &                    &                                     &(day.month.year hr:min)&   \\

\hline
\multirow{2}{*}{NS1} & {\Nu} & 80001041002&07.03.2016 21:01 & 43.3 \\
                      & {\sw} & 00031921100&08.03.2016 12:36 & 2.2 \\
\multirow{2}{*}{NS2} & {\Nu} & 80202014002&12.03.2016 14:18 & 20.2\\
                      & {\sw} & 00031921104&12.03.2016 13:53 & 2.0 \\
\multirow{2}{*}{NS3} & {\Nu} & 80202014004&14.03.2016 19:26 & 20.7\\
                      & {\sw} & 00031921106&14.03.2016 21:39 & 1.0 \\
                  N4 & {\Nu} & 80202014006&29.03.2016 02:41 & 48.4\\         
\multirow{2}{*}{N5} & {\Nu} & 80202015002&30.04.2016 11:26 & 39.1\\
                     & {\sw} & 00081917001$^*$&30.04.2016 21:21 & 1.9 \\
\multirow{2}{*}{NS6} & {\Nu} & 80202015004&26.05.2016 15:51 & 36.3\\
                      & {\sw} & 00081917002&26.05.2016 16:07 & 1.7 \\
\hline
\end{tabular}
\begin{flushleft}
{\bf Note:} $^*$The {\sw} observation 00081917001 is excluded in this work (see text for details). 
\end{flushleft}
\end{table*}

\begin{figure*} 
\centering  
\resizebox{1.5\columnwidth}{!}{\rotatebox{0}{\includegraphics[clip]{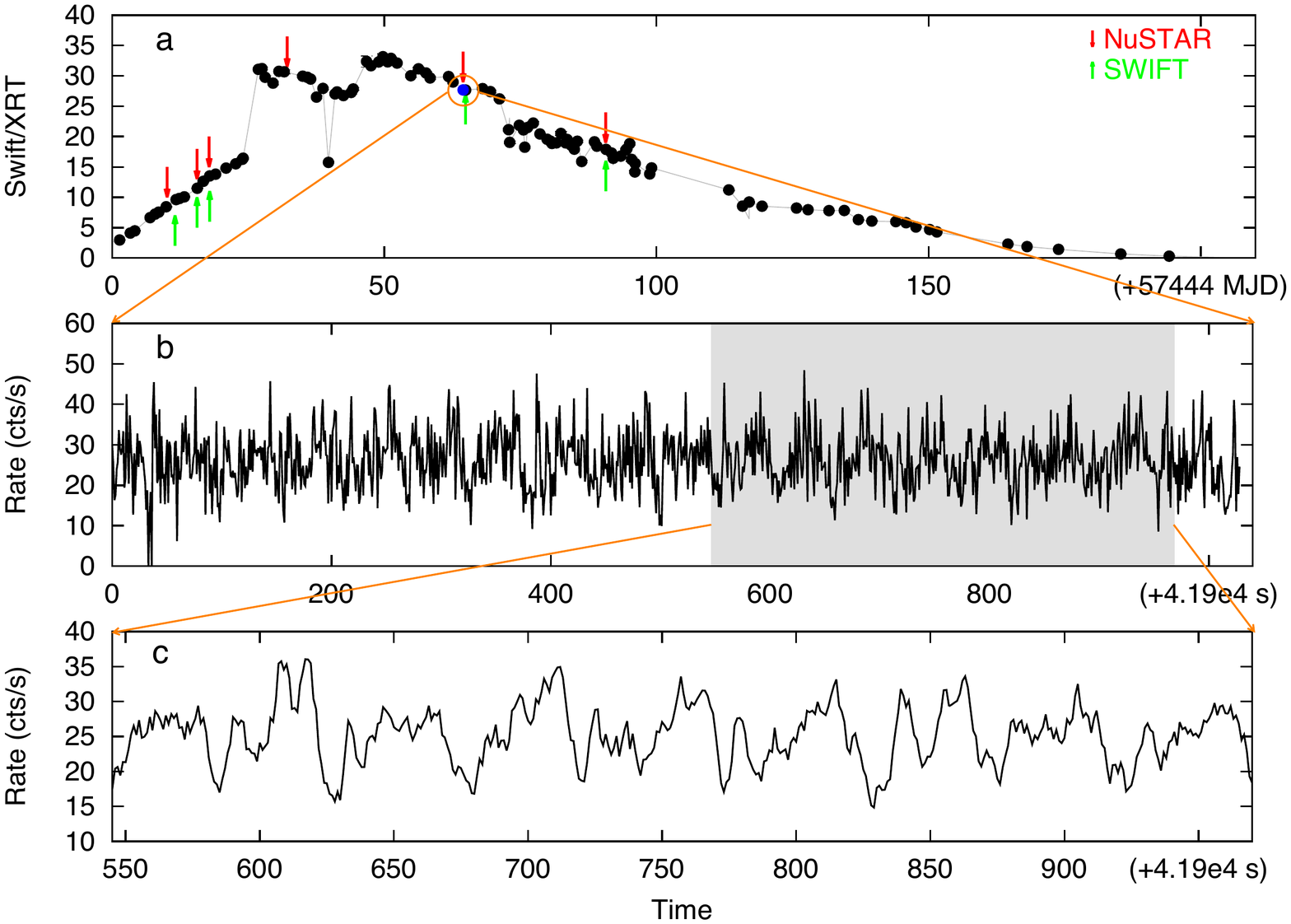}}}
\vspace{-10mm}
\caption{(a) The {\sw}/XRT (0.3$-$10~keV) long-term light curve of {\IGR}.
This light curve is binned at one point per observation.
The red and the green arrows indicate, respectively, the times of the six {\Nu} and the five {\sw} observations listed in Table~1. The blue dot represents the time of the {\Nu} observation N5 displaying the heartbeat variability. (b) Part of the full band {\Nu} light curve of observation N5 at 1-s time resolution. (c) A further zoom in of the light curve of observation N5 smoothed with a square window of 4-s.}
\end{figure*}

\section{results}
\subsection{OUTBURST EVOLUTION}
As it is apparent from the long-term light curve in Fig.~1a, the intensity of {\IGR} increased linearly by a factor of $\sim5$ in the first 23~days (from MJDs 55445 to 57468); at that point, and within 2~days, the intensity increased rapidly by a factor of $\sim2$ and reached a local peak. In the following 26~days the light curve showed small fluctuations in intensity, ranging between 26$-$33~cts/s, and after that the source reached the absolute maximum intensity on MJD 57493.
Since that date, the intensity of the source decayed for 154~days, when the source went back into quiescence.
Compared with the previous outbursts of {\IGR} (e.g., \citealt*{Pahari2014}), the changes in the light curve suggest that the source evolved from the hard, to the soft and back to the hard spectral states. 

On MJD~57508 (observation N5 in Table~1), during the decay of the outburst, {\IGR} displayed heartbeat variability. Fig.~1b and 1c show a part of the {\Nu} light curve of observation N5.

In order to study the spectral evolution of {\IGR} during the 2016 outburst, and to investigate specifically the heartbeat oscillations of observation N5, we first fitted the {\sw} and {\Nu} average spectra jointly, and then carried out phase-resolved spectroscopic study to observation N5.

\subsection{AVERAGE SPECTRA} \label{sec:ave}
We initially fitted the {\Nu} and {\sw} spectra with a model consisting of a disc component, {\sc diskbb} \citep{Mitsuda1984}, and a power-law with a high-energy cut off, {\sc cutoffpl}, to describe the soft and the hard components, respectively. 
A multiplicative term was added to the model to account for calibration uncertainties between {\Nu} and {\sw}; this factor was kept fixed at the value 1 for the {\Nu} data and left free to vary for the {\sw} data during the fits. 
In order to highlight possible reflection features, we only fitted the data over the energy range 0.8$-$5~keV and 10$-$75~keV. As we only used the {\sw} spectra below 3~keV, where reflection is not significant, we do not show the {\sw} data in Fig.~2.
Fig.~2a shows the residuals in terms of sigma with respect to the model {\sc phabs*(diskbb+cutoffpl)}.
We found that the column density of the {\sc phabs} component in observation N5 is not consistent with that in other observations.
Since this is the observation with the heartbeat oscillations, 
and possible changes of the power-law index during the oscillations may induce artificial variations of the best-fitting column density (see e.g. \citealt{2017Cheng}), we excluded the {\sw} data in the analysis of this observation.

The fit with the model above yields prominent positive residuals at around 5 to 10~keV, likely due to an iron emission line, and positive residuals at around 20 to 30~keV, possibly corresponding to a reflection hump (see Fig.~2a).
After we added a {\sc gaussian} component at $\sim$6.5~keV, restricted to be between 6.4 and 7~keV, to fit part of these residuals, we still found some negative residuals at around 7~keV that could be due to absorption by Fe~XXVI. 
In order to test whether the absorption structure is an artefact due to a non-solar abundance in the interstellar material along the line of sight, instead of the {\sc phabs} component we used {\sc vphabs} with the iron abundance left free, the reason being that the $K_{\alpha}$ edge of neutral iron is at 7.1~keV. If we do this, the absorption feature at 7~keV becomes narrower, but does not disappear. 
We therefore added a {\sc gabs} component to represent an absorption line, with the energy and width of the line linked across the observations, and the normalisation free. 
The reduced $\chi^2$, $\chi^2_{\nu}=1.028$ for 7950 degrees of freedom (d.o.f.), and the plot of the residuals indicate that the fit is statistically acceptable (Fig.~2b). 
The total column density and the iron abundance in {\sc vphabs} are covariant in this model, which means that we cannot constrain both simultaneously. 
The energy of the {\sc gaussian} component of all spectra always pegs at 6.4~keV, which indicates that the emission line is likely not symmetric, the accretion disc is lowly ionized, or both.

\begin{figure} 
\centering
\resizebox{1\columnwidth}{!}{\rotatebox{270}{\includegraphics[clip]{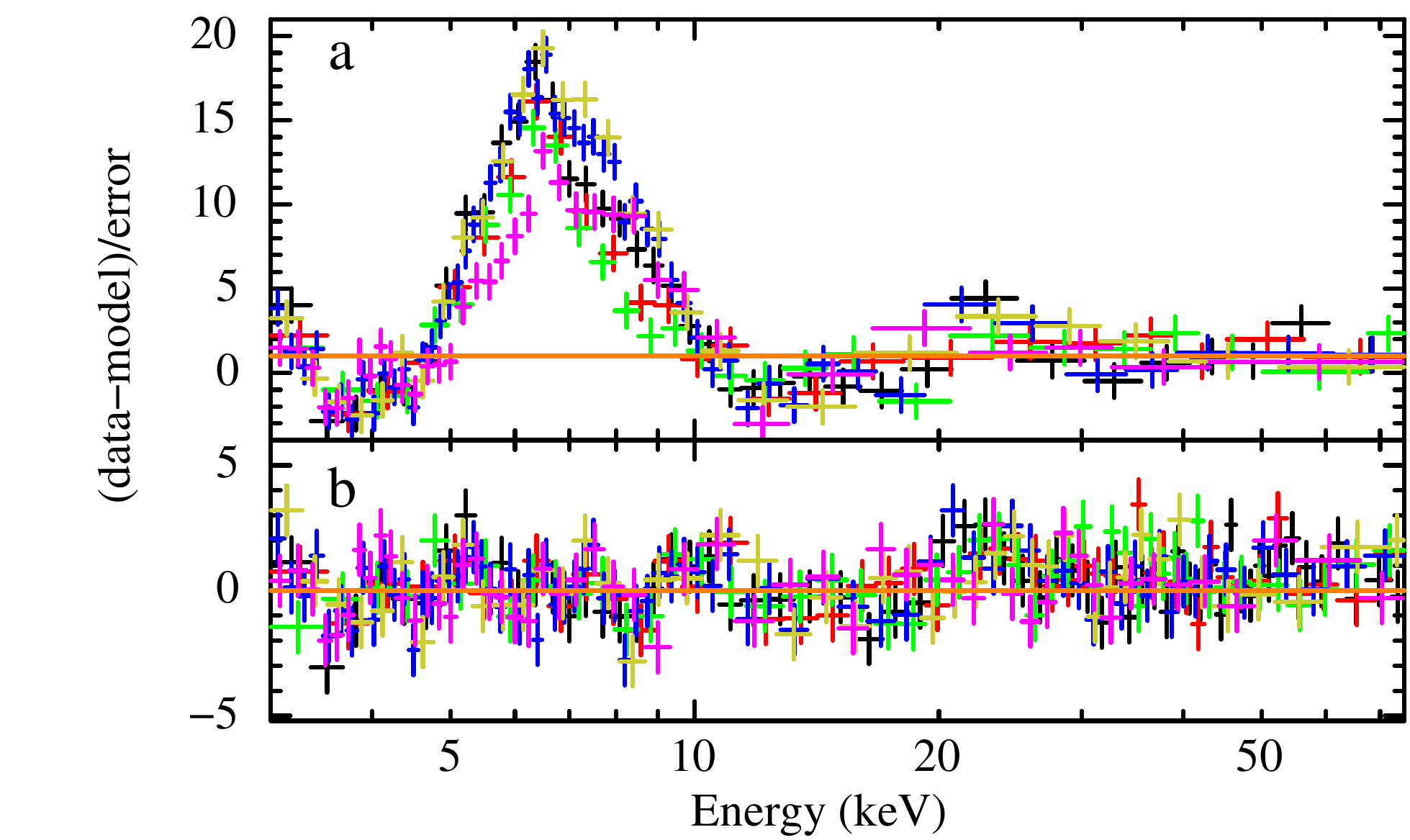}}}
\caption{Residuals in terms of sigmas of different models for the six average {\Nu} spectra of {\IGR}.
(a) Residuals of the model {\sc phabs*(diskbb+cutoffpl)} fitted only over the energies 3$-$5~keV and 10$-$75~keV. 
(b) Residuals of the model {\sc vphabs*(diskbb+gaussian+cutoffpl)*gabs} fitted over the energy range 3$-$75~keV. 
Each colour corresponds to one observation in Table~1, in the sequence black, red, green, blue, olive and magenta. The data have been re-binned for plotting purposes.}
\end{figure}

The broad Fe emission line, plus the possible Compton hump at around 20$-$30~keV in Fig.~2 indicate that (relativistic) reflection may play an important role in the spectra of this source. We hence replaced the {\sc gaussian} and the {\sc cutoffpl} components by the self-consistent relativistic reflection model {\sc relxill} version 0.5b \citep{Dauser2013,Garc2014}. 
In this model, we fixed the outer radius of the disc, $R_{\rm out}$, and the redshift to the source, $z$, to 400~$R_g$ ($R\rm_g$ = $GM/c^2$) and 0, respectively and adopted the spin parameter, $a_{*}=0.998$ (We also tried other options for $a_{*}$; see below). 
We assumed that the inclination of the system, $i$, and the iron abundance, $A\rm_{Fe}$, of the accretion disc do not vary on the time scale of the observations and hence we linked these parameters during the fits to be the same in all observations.
We further linked the inner and outer emissivity indices, $q\rm_{in}$ and $q\rm_{out}$, to be the same within each observation, but left them free to vary between different observations. We found that the changes of the $q\rm_{in}$ were consistent within errors from NS1 to NS6 (see Table~1), therefore, just as with the inclination and the iron abundance, we linked the $q\rm_{in}$ across all the observations.
Once we got a good fit, we fixed the reflection fraction (refl\_frac) to its negative value to use the {\sc relxill} component as a reflection component only and we added a {\sc cutoffpl} to represent the hard component in our data. We did this to be able to get parameters of the refection component separately from the direct emission.
The model then was {\sc const*phabs*(diskbb+relxill+cutoffpl)*gabs}. 
We tried leaving the spin parameter free in our fits (linked across observations), but we could not constrain the spin parameter and the inner radius at the same time.  We also tried fixing $a_{*}$ at $-0.998$, but the best-fitting parameters changed within errors. 

Replacing the {\sc gaussian} component by the {\sc relxill} component improved the fit, such that the $\chi^2$ decreased by $\Delta \chi^2=105.6$ for 3 fewer d.o.f.. 
We show the {\sw} and {\Nu} residuals of each spectrum individually in Fig.~3.
The best-fitting value of the column density, $N\rm_H$, was $1.50\rm~\pm~0.02~\times~10^{22}~cm^{-2}$, and the multiplicative term for the XRT was $0.88\pm0.01$. 
The energy and the width of the {\sc gabs} component were $7.14\pm0.04$~keV and $0.24\pm0.05$~keV, respectively. 
The values of the three linked parameters of the {\sc relxill} component, $q\rm_{in}$, $A\rm_{Fe}$ and $i$, were $3.7\pm0.3$, $3.5\pm0.3$ and $45\degree.3\pm0\degree.7$, respectively. The plots of the $\Delta \chi^2$ of the fit versus those three parameters are shown in Fig.~4. We used the Monte Carlo Markov Chain (MCMC) method to compute the probability distribution function of each parameter, and we obtained all the fluxes from the command {\it flux} in XSPEC, with the errors calculated from the probability distribution function of the flux of each component.

\begin{figure} 
\centering
\resizebox{1\columnwidth}{!}{\rotatebox{270}{\includegraphics[clip]{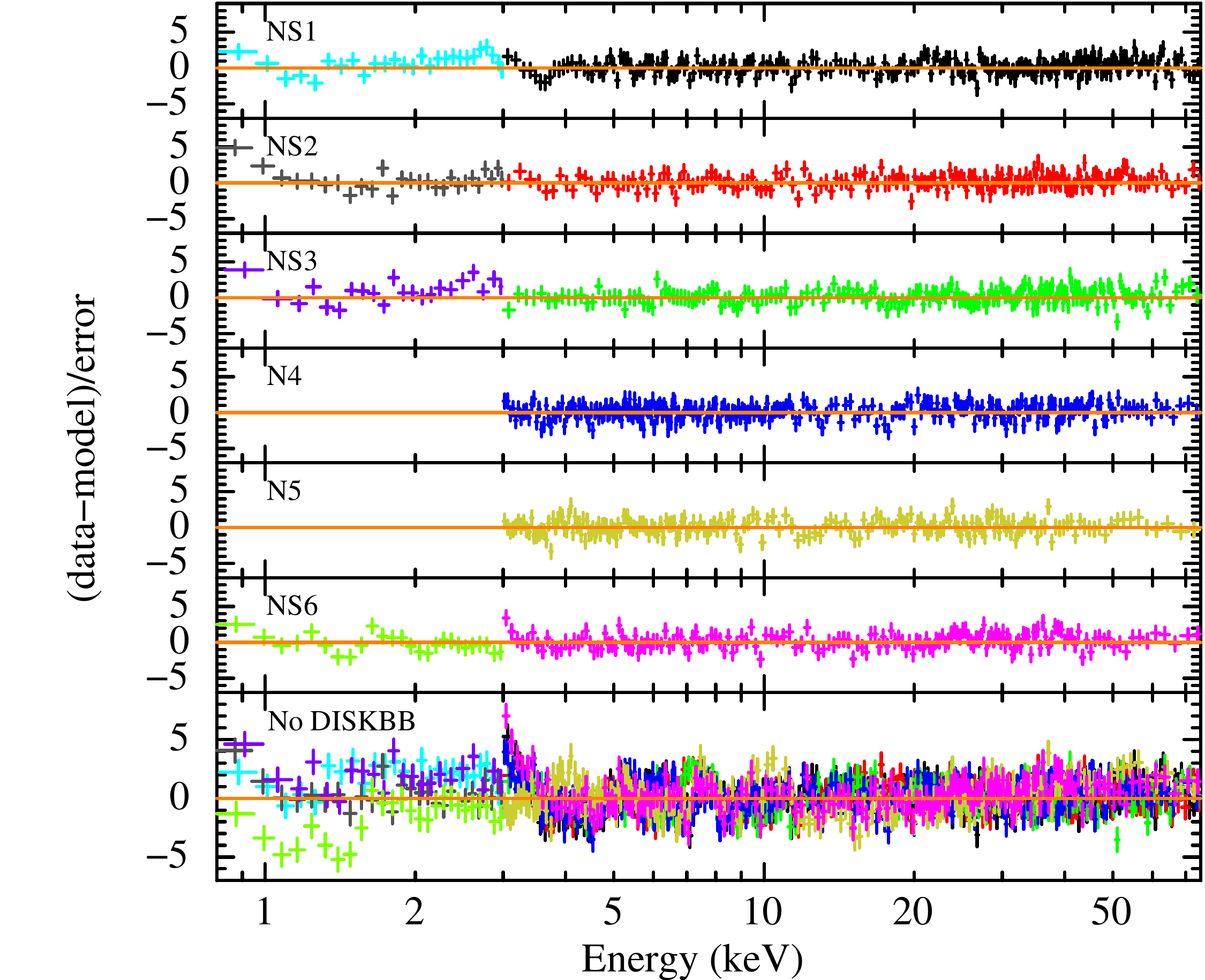}}}
\caption{ Residuals in terms of sigmas of the joint fits to the six {\Nu} and four {\sw} observations of {\IGR}. The first six panels: residuals of the fits with the model {\sc phabs*(diskbb+relxill+cutoffpl)*gabs} of the average spectra of the individual observations; the bottom panel: residuals of the fits with model {\sc phabs*(relxill+cutoffpl)*gabs} of the average spectra of all the observations.}
\end{figure}

Except for the parameters mentioned above, we show the other best-fitting parameters of this model and the flux of the different components versus time in Figs.~5 and 6, respectively. 
The gray areas in Figs.~5 and 6 indicate the three observations (NS1, NS2 and NS3) of {\IGR} in the rising state of the outburst, corresponding to the data used by \cite{Xu2017}. We listed all the best-fitting parameters of this model in Table~A1 in the Appendix.

As it is apparent from Fig.~5, the temperature and the normalisation of the {\sc diskbb} component are anti-correlated with each other.
The photon index, $\Gamma$, and the cut-off energy, $E_{\rm cut}$, of the {\sc cutoffpl} and {\sc relxill} components are anti-correlated as well. Similar to $\Gamma$, the normalisation of the {\sc cutoffpl}, N$_{pl}$, first increases and then decreases with time. The spectrum of observation N5 is insensitive to the value of $E_{\rm cut}$, which pegs at the maximum value of 1000~keV. 
This value is much higher than the upper bound of the {\Nu} energy range, so we fixed $E\rm_{cut}$ in this observation at 1000~keV. The red arrow in panel 4 of Fig.~5 indicates this lower limit of $E\rm_{cut}$.
The inner radius, $R_{\rm in}$, of the {\sc relxill} component followed the same trend as the normalisation of the {\sc diskbb} component. The changes of the reflection fraction, refl\_frac, are opposite to those of the inner radius $R_{\rm in}$.
The ionization parameter, $\xi$, drops first from NS1 to NS3, then increases and remains more or less constant after N4. 
The optical depth\footnote{The optical depth at the line centre is calculated as, $\tau=\rm par3/par2/2\pi$, where par3 and par2 are the strength and $\sigma \rm_{gabs}$ of the {\sc gabs} component, respectively. For more details, see https://heasarc.gsfc.nasa.gov/xanadu/xspec/manual/node236.html.}, $\tau$, of the 7-keV absorption line is consistent with being constant within errors.
As we can see from Fig.~6, the flux of the {\sc diskbb} component first increases, reaches a maximum value at N5, and then decreases again; the flux of the {\sc relxill} component first increases, peaks at N4 and drops after that; the total flux of the model displays the same trend as the flux of the {\sc cutoffpl} component which dominates the emission. 

Because the disc winds are rarely detected in the hard state, it is puzzling that we observed a stable absorption line in all the spectra. Given that the energy and the width of the absorption line was linked between observations, in order to avoid the possibility that some of those features were an artefact of the way we fitted them, we set all the parameters of {\sc gabs} free to vary between the six observations. We got consistent values of the parameters regardless of whether we linked them or left them free between observations.

In order to explore the origin of this moderately broad absorption feature, as we did at the beginning of this section, we replaced {\sc phabs} by {\sc vphabs}, and we refitted the data with the hydrogen column density and the iron abundance free to vary. The difference with what we did earlier is that we now fitted the full model with relativistic reflection. In this case the best-fitting column density and iron abundance are, respectively, $N\rm_{H}=1.24_{-0.16}^{+0.06}~\times~10^{22}~cm^{-2}$ and $A\rm_{Fe}=2.60\pm0.77$ times solar abundance, but the central energy and width of the {\sc gabs} component, $7.15\pm0.03$~keV and $0.28\pm0.03$~keV, respectively, did not change significantly. This suggests that the 7-keV absorption feature is not due to the interstellar medium (ISM). We will discuss this feature further in Section 4.3.

We also checked whether the soft component is required in all the spectra.
If we exclude the {\sc diskbb} component in each observation individually during the simultaneous fits of all the observations, the $\chi^2$ of the fit increased between $\Delta \chi^2=78.1$ and $\Delta \chi^2=709.0$ for 2 d.o.f. more; we show the residuals of the fits without the {\sc diskbb} component in the bottom panel of Fig.~3.
This shows that a soft component improves the fits of all the observations.

\begin{figure} 
\centering
\resizebox{1\columnwidth}{!}{\rotatebox{0}{\includegraphics[clip]{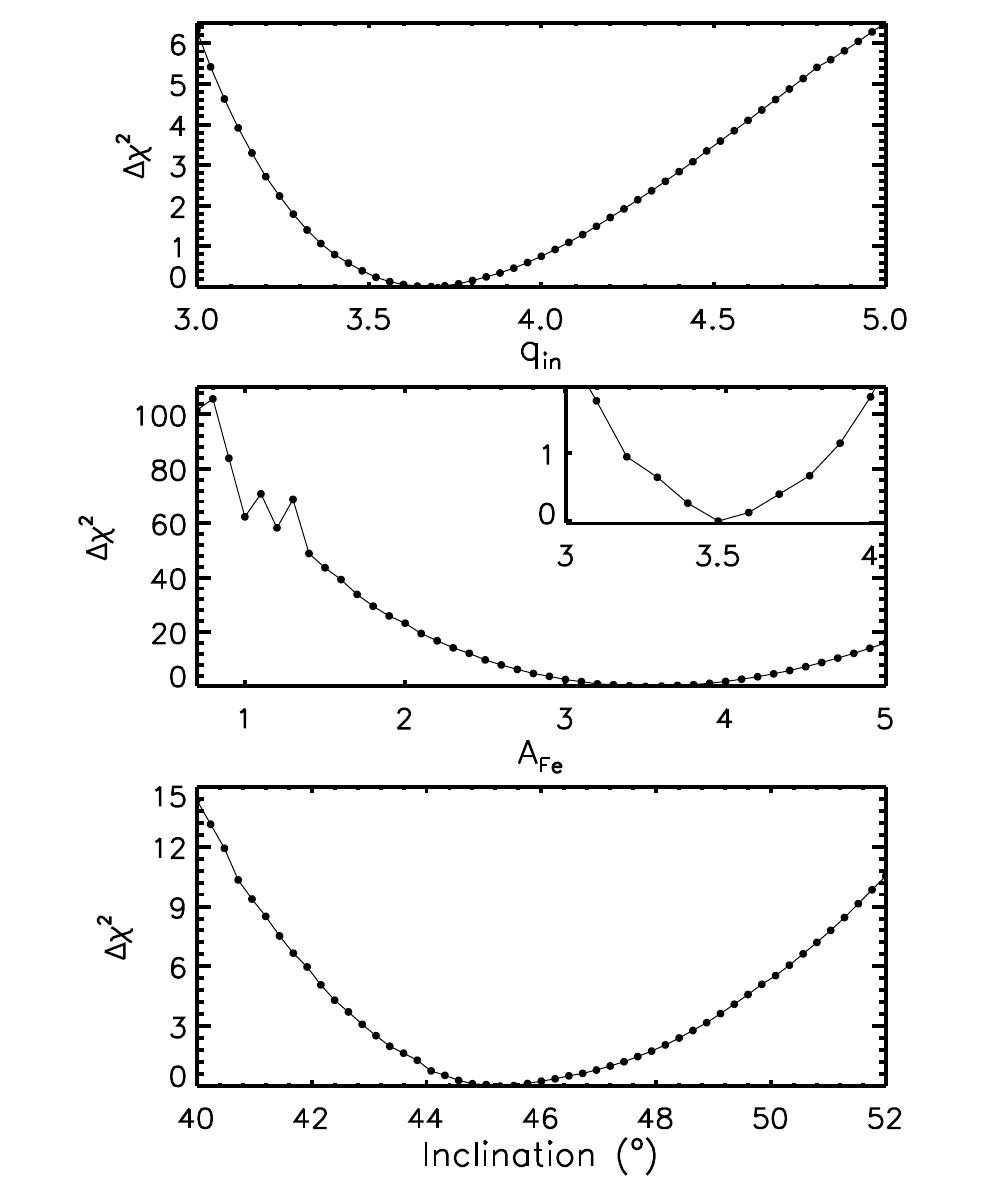}}}
\caption{The change of the goodness-of-fit, $\Delta \chi^2$, versus the inner emissivity index (top panel), the iron abundance in solar units (middle panel) and the inclination angle of the accretion disc (bottom panel) of the {\sc relxill} component in the average spectra for {\IGR}.
The top-right window in the middle panel is a zoom-in of the iron abundance.
The $\Delta \chi^2$ was calculated using the command {\it steppar} in XSPEC over 50 steps in each parameter.}
\end{figure}

\begin{figure} 
\centering
\resizebox{1\columnwidth}{!}{\rotatebox{0}{\includegraphics[clip]{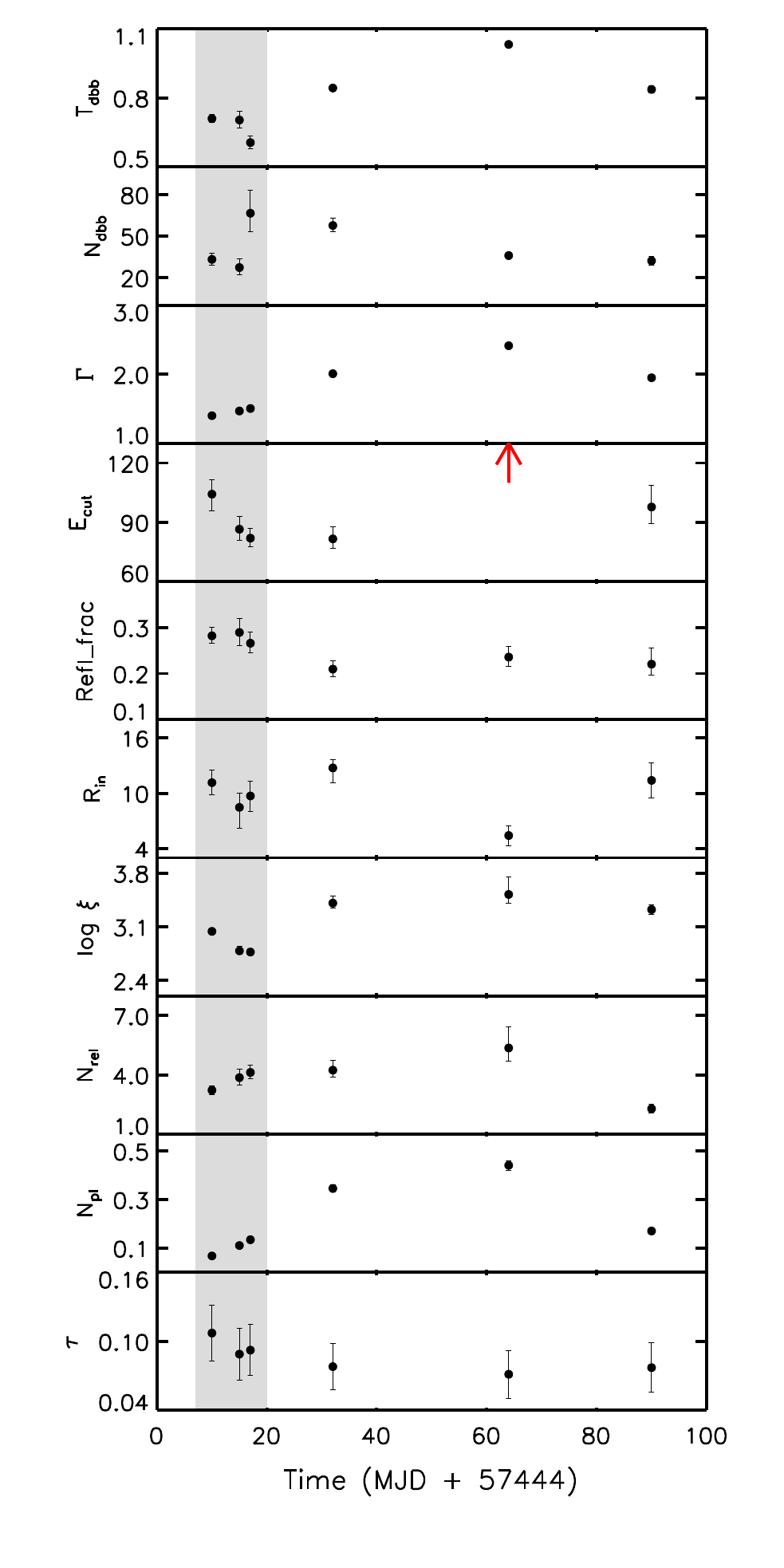}}}
\caption{Evolution of the best-fitting parameters of the average spectra of {\IGR} fitted with the model {\sc const*phabs*(diskbb+relxill+cutoffpl)*gabs}. From the top to the bottom panels the parameters are the disc temperature (keV) and the disc blackbody normalisation ($(R_{\rm dbb}/D_{10})^{2}\cos i$, where $R\rm_{dbb}$ is the the inner disc radius in units of km, $D_{10}$ is the distance to the source in units of 10~kpc and $i$ is the inclination angle of the disc), the photon index and the cut-off energy (keV) of the power-law component, the reflection fraction, the inner disc radius ($R\rm_{g}$), the disc ionisation ($\rm erg~cm~s^{-1}$) and the normalisation of {\sc relxill} component, the normalisation ($\rm photons~cm^{-2}s^{-1}kev^{-1}$ at 1~keV) of the {\sc cutoffpl} component and the optical depth of the absorption line. The red arrow indicates that the $E\rm_{cut}$ of observation N5 pegged at 1000~keV. Errors are quoted at the 1-$\sigma$ confidence level. The parameters in the gray area are from the observations NS1, NS2 and NS3 which were used in \protect\cite{Xu2017}.} 
\end{figure}

\begin{figure} 
\centering
\resizebox{1\columnwidth}{!}{\rotatebox{0}{\includegraphics[clip]{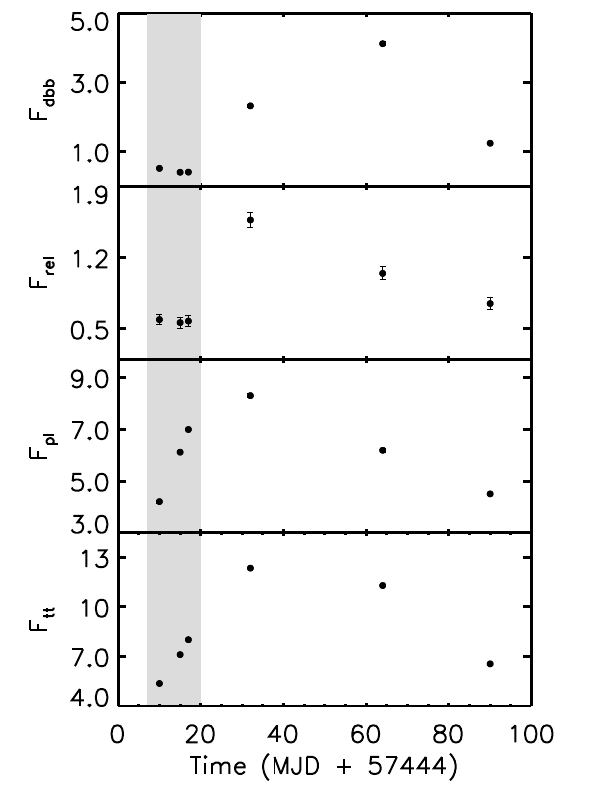}}}
\caption{Evolution of the unabsorbed flux of each component for the fit of the average spectra of {\IGR}.
From the top to the bottom panels, $F\rm_{dbb}$, $F\rm_{rel}$, $F\rm_{pl}$ and $F\rm_{tt}$ represent, respectively, the unabsorbed fluxes of the components {\sc diskbb}, {\sc relxill}, {\sc cutoffpl} and the entire model in the 2--10~keV range in units of $\rm 10^{-10}~erg~cm^{-2}s^{-1}$. Errors are quoted at the 1~$\sigma$ confidence level. The fluxes in the gray area are from the observations NS1, NS2 and NS3 which were used in \protect\cite{Xu2017}.}
\end{figure}

\subsection{PHASE-RESOLVED SPECTRA} \label{sec:phase}
As can be seen from Fig.~1b and 1c, {\IGR} showed heartbeat variability during observation N5. We therefore extracted spectra as a function of the phase of the oscillation to investigate how the spectral components change on the QPO cycle. 

The oscillations within each orbit of observation N5 were irregular and showed small frequency shifts, while the frequency of those oscillations changed significantly between orbits (see top panel in Fig.~7). 
We folded the data of observation N5 using the method described in Appendix~A of \cite{Court2017}. We first used a numerical algorithm to determine the time-coordinate of the peak of each individual heartbeat. We assigned a sequential integer value to each of these (i.e. the first peak was assigned a value of 1, the second a value of 2, etc.) to create a function $\Phi(t)$. We then fit a monotonically increasing univariate cubic spline to the data to obtain a function $\Phi(t)$ valid for all times. The value of $\Phi(t)~~\rm mod~1$ is then the phase at any given time $t$. We then binned the data based on the value $\Phi(t)$ at their $t$-coordinate. Fig.~8 shows the folded waveform. This method assumes that heartbeat flares have high enough amplitude to be distinguished by eye from light curves.

There are 17 orbits in total in observation N5. However, we only included the first 14 orbits in this work because the amplitude of the heartbeat flares in the last three orbits is not high enough to be distinguished by eyes. We then defined the Good Time Intervals (GTIs) of the 14 orbits for each phase bin and we extracted phase-resolved spectra using the command {\it nuproducts}.

In order to test whether the spectra of different orbits can be combined, we first fitted the average spectra of the 14 orbits individually with the model {\sc phabs*(diskbb+powerlaw)}, in the energy of 3$-$5~keV and 10$-$75~keV to avoid the influence of the possible emission line, and found that all the parameters were consistent within errors between the different orbits (Fig.~7).
Therefore, we combined all the spectra at the same phase from different orbits and different instruments (FPMA/B) together to increase the signal to noise ratio at each phase, based on the assumption that the spectral properties are independent of the QPO frequency, but only depend on the phase of the oscillation.

\begin{figure} 
\centering
\resizebox{1\columnwidth}{!}{\rotatebox{0}{\includegraphics[clip]{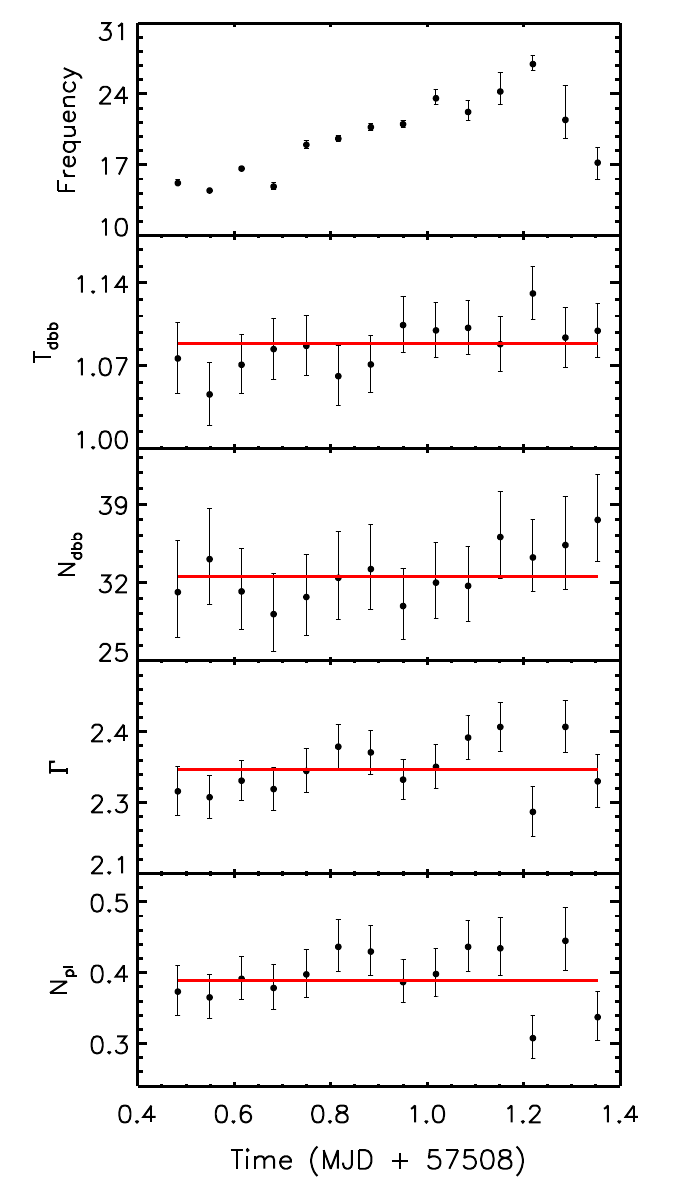}}}
\caption{Evolution of the QPO centroid frequency and the best-fitting parameters of the first 14 orbits of observation N5 of {\IGR} fitted with the model {\sc phabs*(diskbb+powerlaw)}. The parameters from top to bottom are: the QPO frequency (mHz), the disc temperature, the disc blackbody normalisation, the photon index and the power-law normalisation. The parameters in the last four panels are in the same units as in Fig.~5. The red lines in the last four panels, respectively, represent a fit with a constant to each parameter.}
\end{figure}

\begin{figure} 
\centering
\resizebox{1\columnwidth}{!}{\rotatebox{0}{\includegraphics[clip]{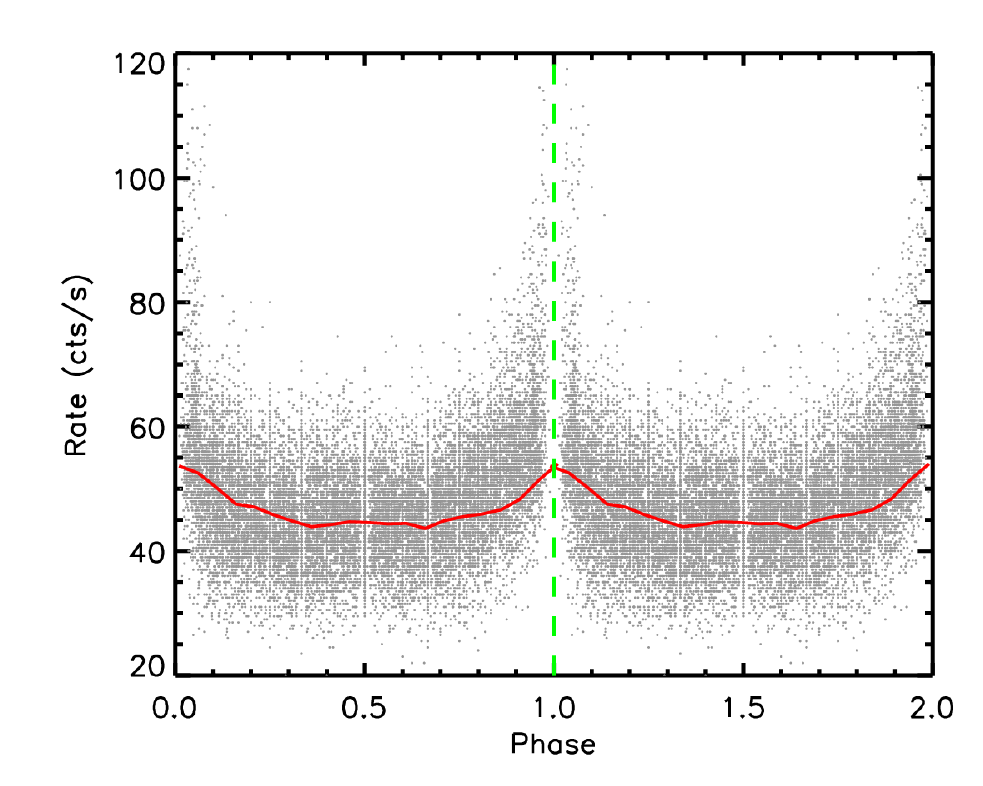}}}
\caption{Folded waveform from the combined FPMA and FPMB light curves of observation N5 of {\IGR} with a time resolution of 1-s. The red solid line represent the average waveform.}
\end{figure}

We initially extracted spectra in ten phases.
In Fig.~9a, we show the deviations of the spectra at phases 0, 0.5, 0.7, 0.8, with respect to the average spectrum of the full observation N5. Following the same procedure as in Section~3.1, we fitted the ten spectra simultaneously with the model {\sc phabs*(diskbb+cutoffpl)} over the energy ranges 3$-$5~keV and 10$-$75~keV, with the column density fixed at $N\rm_{H}=1.50~\times~10^{22}~cm^{-2}$, derived from the fits to the average spectra.
As in the analysis described in Section~3.2, we got broad residuals at 5$-$10~keV and 20$-$30~keV (see Fig.~9b). 
After adding the reflection component, {\sc relxill}, to the model, an absorption feature was still visible at around 7~keV in the spectra, so we included a {\sc gabs} component to the model. We show the residuals in terms of sigma for this model in Fig.~9c, for the same phases as in Fig.~9a.
Due to the low count rate of {\IGR}, and the limited exposure time of each phase-resolved spectrum, the data were not good enough to constrain all the parameters of {\sc relxill} simultaneously. Therefore, we reduced the numbers of phase bins to six to further enhance the signal to noise ratio of each spectrum. 

\begin{figure}
\centering
\resizebox{1\columnwidth}{!}{\rotatebox{270}{\includegraphics[clip]{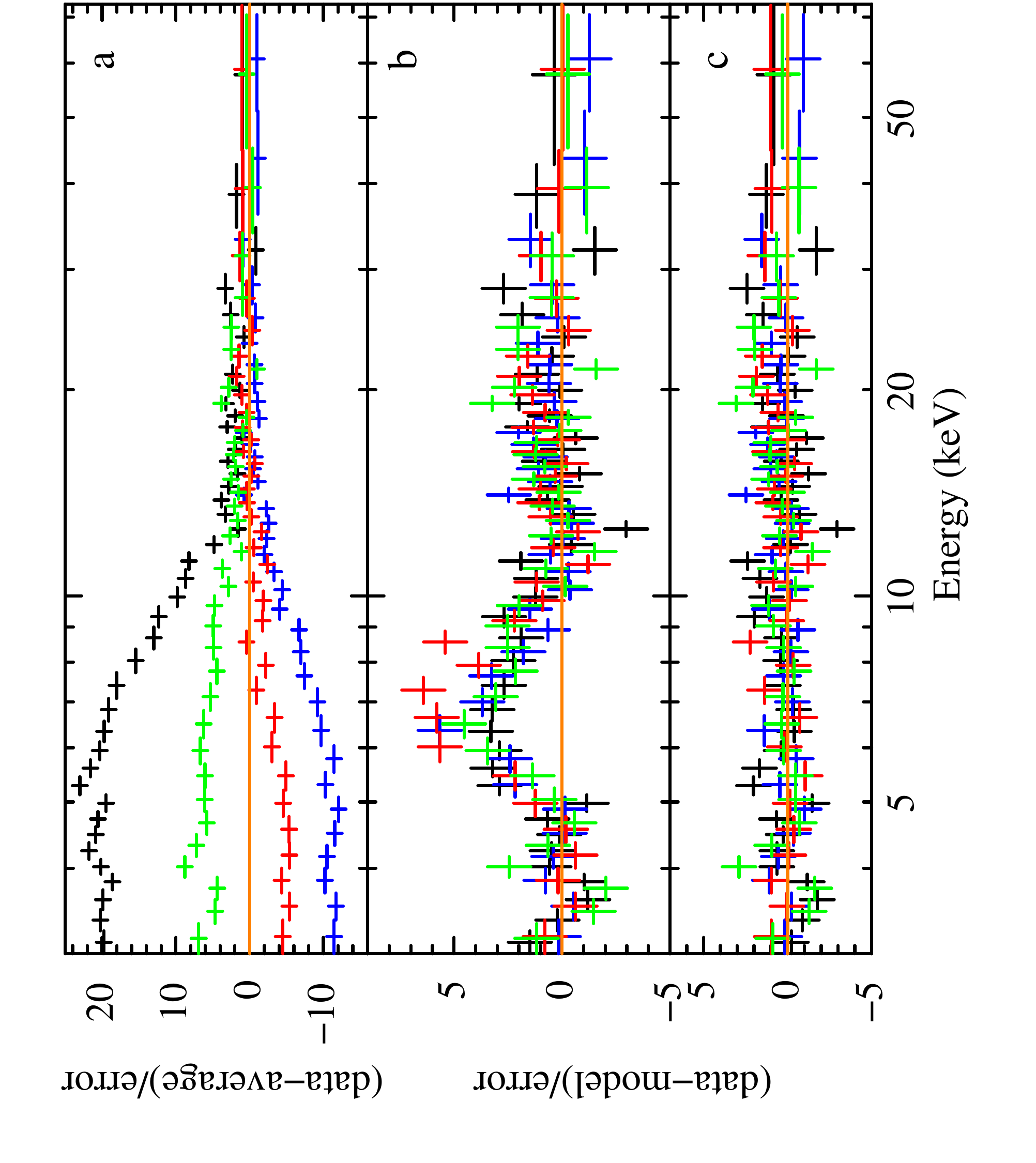}}}
\caption{(a) Deviations from the average spectrum of observation N5 of {\IGR} at four phases. The black, red, green and blue lines indicate the deviations from the average spectrum at, respectively, phases 0, 0.5, 0.7 and 0.8. (b) The residuals in terms of sigma for the model {\sc phabs*(diskbb+cutoffpl)} fitted over the energy ranges 3$-$5 and 10$-$75~keV for the same four phases as shown in panel~a. (c) the residuals in terms of sigma for the model {\sc phabs*(diskbb+relxill+cutoffpl)*gabs} fitted over the energy range 3$-$75~keV for the same four phases as shown in panel~a.}
\end{figure}

For the parameters in the {\sc relxill} component, as in Section~3.2, we first fixed the spin parameter, $a_{*} = 0.998$, the outer radius of the disc, $R\rm_{out}$ = $400~R\rm_g$ and the redshift to the source, $z = 0$, and set the inclination angle of the system to $i=45\degree.3$ and the iron abundance to $A\rm_{in}=3.5$, corresponding to the best-fitting parameters of the average fits. If we let the inner radius of the disc, $R\rm_{in}$, free across phases, the best-fitting values had very large uncertainties; therefore we decided to link $R\rm_{in}$ across phases. 

We used an F-test to test how significant the change of each {\sc relxill} parameter was. For this, we calculated the F-statistic from the two $\chi^2$ values when we either let the parameters free to change or linked them between phases. In order to reduce the effect of the parameters with large uncertainties on the fit, if the significance of a parameter was less than 1~$\sigma$, we fixed that parameter at the value derived from the fit to the average spectrum. 
Following this criterion, in the end we fixed the emissivity index, $q_{in}$, the ionization of the disc, $\log \xi$, and the inner radius of the disc, $R\rm_{in}$, of the {\sc relxill} component at their best-fitting values for the fits to the average spectrum of observation N5, and the cut-off energy $E\rm_{cut}$ at 1000~keV. The other parameters, relf\_frac and $\Gamma$, were free to vary in the fit (see all the values in Table~2A).
Because we are interested in the normalisation of the {\sc relxill} component, even though the significance of the variability of this parameter was only $0.8~\sigma$, we did not link this parameter across phases.
By adding a {\sc gabs} component at 7~keV to our model, the $\chi^2$ decreased by $\Delta \chi^2=38.4$ for 3 d.o.f. fewer. We therefore included this component in the fits, with the energy, width and normalisation of the {\sc gabs} component linked between phases.

The reduced $\chi^2$ of this model was $\chi^2_{\nu}=1.02$ for 3512 d.o.f.. 
We show the best-fitting parameters of the model that were free versus phase in Fig.~9 and we give all the best-fitting parameters in Table~A2 in the Appendix.
As we can see from Fig.~10, except the {\sc diskbb} normalisation and the refl\_frac (panels~c and e), the other parameters display a similar trend with phase as the net count rate. 
The reflection normalisation, $N\rm_{ref}$ (panel~f), appears to be correlated with the power-law normalisation, $N\rm_{pl}$ (panel~g), but the F-test indicates that $N\rm_{ref}$ does not change significantly.

We also show the unabsorbed flux of each component and of the entire model versus phase in Fig.~10 (panels h to k). 
The variation of the reflection flux (panel~i) is consistent with errors. Although the flux of the soft, {\sc diskbb}, component is smaller than that of the hard, {\sc cutoffpl}, component (panels~h and j), the variation of the {\sc diskbb} temperature is more significant than that of the {\sc cutoffpl} photon index (panels~b and d).

\begin{figure}
\centering
\resizebox{1\columnwidth}{!}{\rotatebox{0}{\includegraphics[clip]{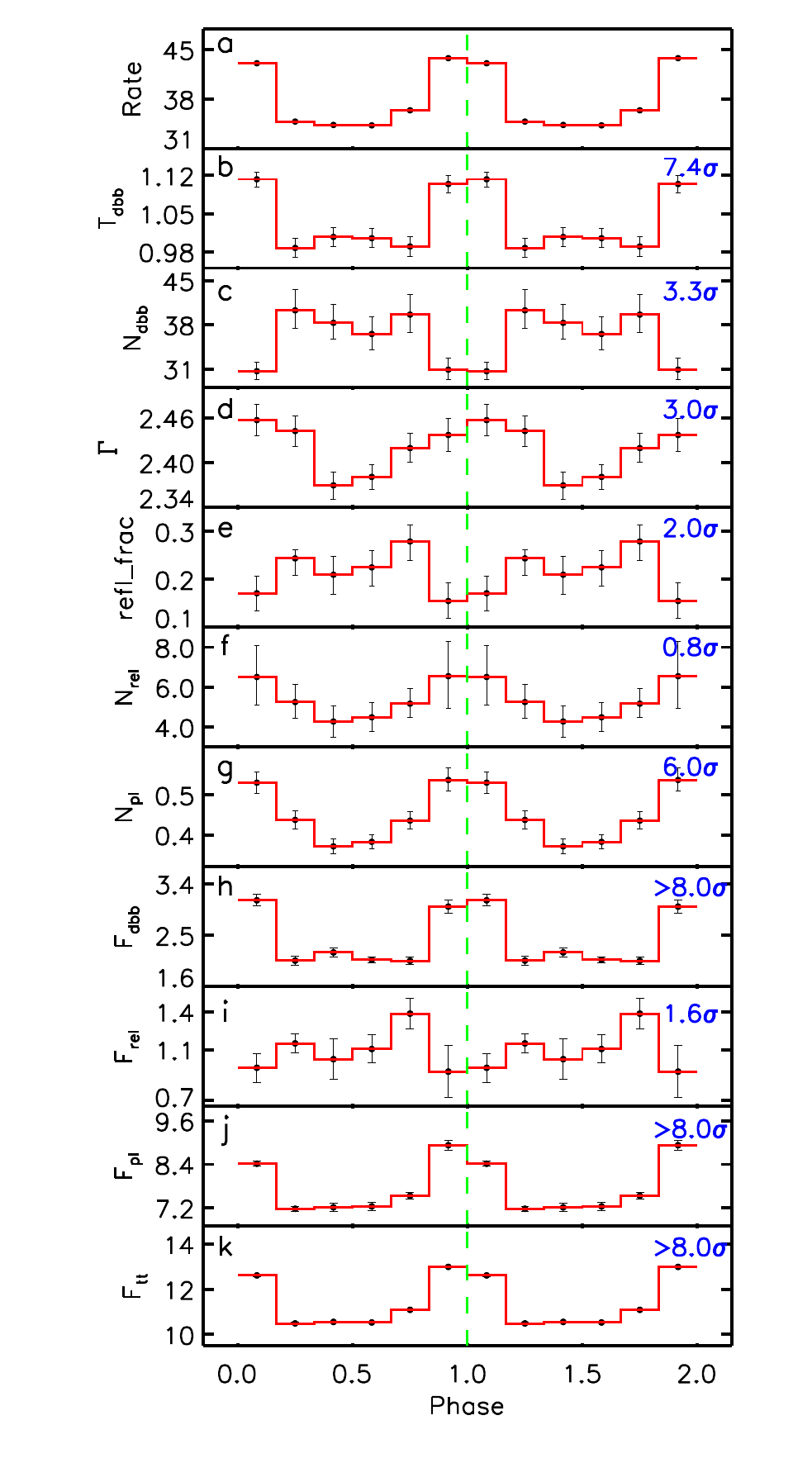}}}
\caption{Changes of best-fitting parameters of the {\Nu} phase-resolved spectra of observation N5 of {\IGR} fitted with the model {\sc phabs*(diskbb+relxill+cutoffpl)*gabs}. Two cycles of the oscillation are plotted here for clarity. From panel a to k we plot: net count rate (cts/s) in the 3$-$75~keV band,
{\sc diskbb} temperature and normalisation, power-law photon index, refl\_frac and normalisation of the {\sc relxill} component ($10^{-3}$), normalisation of the {\sc cutoffpl} component and fluxes of each component in this model. All the fluxes are unabsorbed, in the 3$-$75~keV band. 
All the parameters are in the same units as in Figs.~5 and 6.
At the top-right corner of panels~b to g we give the significance of the variation of the parameters when they are free to vary compared to when they are linked across phases.}
\end{figure}

\section{Discussion}
We analysed six {\Nu} and four {\sw} spectra obtained during the 2016 outburst of the black-hole candidate and transient source {\IGR}.  
The presence of a broad emission line at $\sim6.4$~keV and an emission hump at 20$-$30~keV in the spectrum of this source indicates that reflection off the accretion disc is important in this system. Fits to the average spectra with a reflection model yield an inclination angle of $\sim 45\degree$ and a relatively high iron abundance of $\sim3.5$ times the solar abundance.
One of the observations was carried out at the time that {\IGR} displayed the so-called heartbeat variability. Here we fitted, for the first time, the reflection spectrum of {\IGR} as a function of the phase of the heartbeat QPO and found that the variation of the reflection component is not correlated with that of the direct component.
We discuss the spectral evolution of this source in different outbursts first, and then compare the phase-resolved spectroscopy of {\IGR} with that of {\rgs}.

\subsection{Comparison between the 2011 and the 2016 outbursts of {\IGR}}

\cite{Pahari2014} studied the X-ray timing and spectral evolution of {\IGR} during the first 66 days of the 2011 outburst of this source using {\it RXTE}/PCA and {\sw}/XRT observations. They fitted the data with a model that consisted either of a {\sc powerlaw} or a {\sc diskbb+powerlaw}, depending on the state of the source. The total 2$-$60~keV unabsorbed flux for the combined PCA and XRT spectra was in the range 3.1$-$18.3~$\rm \times~10^{-10}~erg~cm^{-2}~s^{-1}$ when the source was in the hard/soft intermediate states, from 2011 February 23 to 2011 March 25.
The total 2$-$60~keV unabsorbed flux for the six average spectra that we analysed here varies between 11 and 24~$\rm \times~10^{-10}~erg~cm^{-2}~s^{-1}$, which shows that the outbursts of the source in 2011 and 2016 reached comparable flux levels. The main difference in the spectra of those two outbursts is the existence of a reflection component in the 2016 outburst, which does not appear to be required in the 2011 outburst.

The long-term XRT light curve of {\IGR} shown in Fig.~1a suggests that in the 2016 outburst the source evolved from the hard to the soft and back to the hard spectral state. The relatively flat power law and the low disc temperature suggest that, in the first three observations, NS1 to NS3 (see Fig.~5), the source was in the hard state. The low disc flux and the increasing power-law and total fluxes from NS1 to NS3 (see Fig.~6) also support this idea. The power-law index increases in the following observation, N4. The power-law and total fluxes keep increasing from NS3 to N4, while the disc flux increases as well. 
Those changes indicate that a hard-to-soft state transition takes place, suggesting that the source at that time was either in the hard intermediate or the soft state.
The disc flux contribution in N4 is less than 19\% of the total flux, which argues against the possibility of the source having reached the soft state (see the definition of X-ray spectral states in \citealt{Remillard2006}).
The disc temperature and the power-law photon index reach a maximum value in N5. The cut-off energy, on the other hand, becomes very large in this observation (we discuss this later in this section). The disc flux reaches its maximum value in observation N5, while the power-law and total fluxes decrease from N4 to N5. The contribution of the disc to the total flux increases up to 37\% in observation N5. 
Compared with the the best-fitting parameters listed in Table~2 of \cite{Pahari2014}, the source in N5 was most likely in the soft intermediate state; due to the presence of heartbeat oscillations in this observation, here we call this state the heartbeat state. 
In NS6, the drop of the disc temperature, the photon index, the disc flux and the power-law and total fluxes suggests that the source evolved back to the hard state. 

Using data from {\it INTEGRAL} and {\it SWIFT}, \cite{Capitanio2012} studied the spectral evolution of {\IGR} during the first two months of the 2011 outburst and found that
at the end of the transition to the soft state the high-energy cut-off became undetectable up to 200~keV. Similarly, the high-energy cut-off in observation N5 also became undetectable in our work.
\citealt*{Motta2009} investigated the evolution of the high-energy cut-off in the X-ray spectrum of GX~339--4 across a hard-to-soft transition. The change of the cut-off energy from NS1 to N5 in our work followed a trend similar to that in \cite{Motta2009}. They explained the change of the cut-off energy in the intermediate state transition as possibly due to a jet,
because these changes took place very close in time to the moment in which the jet emission appeared.
Even though \cite{Rodriguez2011} reported a radio flare in the soft intermediate state of {\IGR} during the 2011 outburst, due to the lack of radio observations, we cannot verify the appearance of a jet here.

\subsection{The continuum in {\IGR}}
\cite{Xu2017} reported a spectral and timing study of three {\Nu} and {\sw} observations of {\IGR} in the hard state of the 2016 outburst, which are observations NS1, NS2 and NS3 in our work. They found reflection features in all the three {\Nu} spectra and fitted the {\Nu} and {\sw} spectra jointly with four self-consistent reflection models with a thermally Comptonised continuum described by {\sc nthcomp} (\citealt*{Zdziarski1996}; \citealt{Zycki1999}). 

Neither a soft component nor an absorption line at $\sim7$ keV have been reported by \cite{Xu2017}. 
However, as we showed in Section~3.2, including a soft component improves the fits significantly, even though it only contributes 5\% to 9\% to the total flux in the 2--10~keV energy range when the source was in the hard state (see Table~A1). Regarding the {\sc gabs} component, if we deleted this component in each observation individually during the simultaneous fits of all observations, the $\chi^2$ increased between $\Delta\chi^2= 20.2$ and $\Delta\chi^2=39.6$ for 1 d.o.f. more. We therefore conclude that the absorption line is also required by the fit.

We later compared the fits using either {\sc cutoffpl} or {\sc nthcompt} to the spectra of observation NS1, the hardest state in this outburst. We fitted NS1 with the simplified models {\sc phabs*(diskbb+gauss+cutoffpl)*gabs} and {\sc phabs*(diskbb+gauss+nthcomp)*gabs} and got, respectively, $\chi^2=2876.9$ and $\chi^2=2904.1$, both for 2694 d.o.f.. Even though both fits are reasonable, the electron temperature in {\sc nthcompt}, $kT\rm_{e}=18.7\pm0.4$~keV, is more than three times lower than the cut-off energy in {\sc cutoffpl}, $E\rm_{cut}=63.8\pm2.7$~keV, but the photon index in the latter, $\Gamma=1.57\pm0.01$, is larger than that in the former, $\Gamma=1.18\pm0.02$.
However, despite how different the values of the parameters in these two models are, the relative evolution of the power-law parameters versus the source flux in the 2$-$10~keV range are consistent with being the same in both cases.

\subsection{Absorption features in {\IGR}}

Ionised disc winds have been detected in the soft spectral state in one {\it Chandra} observation of {\IGR} during the 2011 outburst by \citep{King2012}. They found two absorption lines at 6.91~keV and 7.32~keV that they identified with blue-shifted Fe~{\sc xxv} and Fe~{\sc xxvi}, with velocities of 9,000 and 15,000~$\rm km~s^{-1}$, respectively. If we associate the 7-keV absorption line with H-like iron, Fe~{\sc xxvi}, line at 6.97~keV, this would correspond to a velocity of $7,300\pm1,700\rm~km~s^{-1}$. Although less likely, the 7-keV absorption line could also be associated with a blueshift from the He-like Fe~XXV at 6.7~keV, which would imply an even higher outflow velocity, $19,700\pm1,800\rm~km~s^{-1}$. 
The width of the 7-keV absorption line implies a turbulent velocity of $10,000\pm2,000~\rm km~s^{-1}$. Given the lower resolution of our {\Nu} spectra compared to the {\it Chandra} spectra of \cite{King2012}, we cannot rule out the possibility that line width in our observations is, at least in part, the result of blending of the two components seen in the {\it Chandra} spectra.

A large study of BH LMXBs by \cite{Ponti2012} revealed that disc
winds (in high-inclination systems) are mainly found in the spectrally soft X-ray state, but
\cite{Homan2016} suggested that in LMXBs with luminosities above a few tens of percent of the Eddington luminosity, disc winds can be present in the hard state. 
\cite{King2012} actually analysed two {\it Chandra} observations, one on 2011 August 1 and another one on 2011 October 16, but they only observed absorption features in the latter, which indicated a variable absorption geometry in {\IGR}.
Different from their result, the 7-keV absorption feature appears in all the six observations that we analysed here, and had a strength that is consistent between observations. 
Assuming a distance of 20~kpc and a mass of 3~$\rm M_{\odot}$ BH in {\IGR}(\citealt{Altamirano2011}; see also Section~4.6), the corresponding luminosity during those observations was likely at $\sim$ 5\%--14\% of the Eddington luminosity.
We thus cannot completely exclude the possibility that this apparently stable absorption feature in {\IGR} is from disc winds.

Alternatively, an absorption feature could be produced by the ISM. For instance, emission lines have been claimed in the past that could in fact be an artefact due to non-solar elemental abundance in the ISM (e.g., \citealt{Wang2016}), which triggered us to employ {\sc vphabs} instead of {\sc phabs} to explore the origin of the absorption line.
Our analysis (see Section 3.2), however, shows that the width of this absorption feature is hardly influenced by variations of the iron abundance in the photoelectric absorption component.


\subsection{The reflection component in {\IGR}}

As we can see from Fig.~3, the iron abundance inferred from the reflection model here, $A\rm_{Fe}=3.5\pm0.3$ of the solar abundance, is much higher than the value $\sim0.7$ found by \cite{Xu2017}; notice, however, that \cite{Xu2017} only fitted the spectra of {\IGR} in the hard state. 
In order to test whether the discrepancy comes from the {\sc gabs} component, which was not included by \cite{Xu2017}, we excluded the {\sc gabs} component in our model to fit NS1. Besides the fact that the fit became worse, the $\chi^2$ increased by $\Delta \chi^2=39.2$ for 2 d.o.f. fewer, the iron abundance decreased to $A\rm_{Fe}=1.4\pm0.4$, which is statistically consistent with the value found by \cite{Xu2017}.
Using simple reflection models, \cite{Furst2015} reported a very high iron abundance in the accretion disc of GX~339--4. 
However, when they allowed the photon index of the incident continuum spectrum to differ from that of the observed primary continuum, the fit improved statistically and the iron abundance decreased. 
Although Cyg~X--1 has also been reported as a black-hole binary with a high iron abundance (e.g., \citealt{Parker2015,Walton2016}), \cite{Tomsick2018} recently found that this high value is related to density effects in the reflector. A high-density model improved their fit and, at the same time, yielded an iron abundance consistent with solar.   
Both works suggested that the iron abundance strongly depends on the model assumptions, which appears to be similar with our case.

{\Nu} has been frequently used to estimate inclination angles of the accretion disc in LMXBs from reflection spectra (e.g., \citealt{Furst2015,Ludlam2016,Wang2017}). 
We find that the inclination of the accretion disc of {\IGR} measured from fitting the reflection model is $i\sim45\degree$, which is consistent with the low inclination angle of 30$\degree-$40$\degree$ proposed by \cite{Xu2017}.
When there is no reflection component, one can also estimate the inclination angle from other spectral components. 
For instance, \cite{King2012}, \cite{Capitanio2012} and \cite{Rao2012} analysed data of {\IGR} during its 2011 outburst and obtained a high inclination angle, up to $\sim$$70\degree$.
\cite{Xu2017} speculated that this discrepancy is the result of a warped accretion disc and suggested that the low inclination angle was either the orbital inclination, or somewhere in between the orbital and the inner disc inclination angle.

\subsection{The spin parameter of {\IGR}}
Using simultaneous {\it RXTE} and {\it XMM-Newton} observations, \cite{Rao2012} reported a phase-resolved spectroscopic study of {\IGR} in the heartbeat state during the 2011 outburst of the source. 
By fitting the continuum, they obtained an inclination angle $i\gg50\degree$ and a spin parameter $a_{*}<0.2$, suggesting that {\IGR} has a low, or even negative, spin parameter.

Inspired by this, we re-analysed the effect of the spin parameter to the fitting results. 
If we assume that in all observations of {\IGR} the inner radius of the disc in the {\sc relxill} component, $R\rm_{in}$, is located at the radius of innermost stable circular orbit (ISCO), $R\rm_{ISCO}$, the spin parameter for the average spectra pegs at the lowest possible value, $-0.998$. Using MCMC we find that, with 90\% confidence, the best-fitting spin parameter is between $-0.998$ and $-0.18$. Because the fit with $R_{\rm in}=R_{\rm ISCO}$ is worse than that with $R\rm_{in}$ free, $\chi^2$ increased by $\Delta \chi^2=11.1$ for 3 d.o.f., it is likely that this assumption is incorrect and that the disc is probably truncated in all the observations except, perhaps in observation N5, which has the smallest $R\rm_{in}$, in which the disc may extend down to the ISCO.

We therefore only fixed the value of $R\rm_{in}$ at $R\rm_{ISCO}$ in this observation, leaving $R\rm_{in}$ free in the other ones, and re-fitted the average spectra. In this case we got $a_{*}=0.05_{-0.69}^{+0.16}$ (90\% confidence).
If we set $R\rm_{in}$ = $R\rm_{ISCO}$ in the fits of the phase-resolved spectra of this observation, we get almost the same $\chi^2$ for 1 d.o.f. less, comparing with the fit with $R\rm_{in}$ free in observation N5, which indicates that we cannot discard the possibility that the inner radius of the disc in N5 is at the ISCO. 
The best-fitting spin parameter for the phase-resolved spectra in N5 is $a_{*}=-0.28_{-0.44}^{+0.64}$ (90\% confidence).
From the previous results, the joint probability distribution for the spin parameter indicates that, with 90\% confidence, $-0.13\leq a_{*}\leq0.27$.
Hence, if the assumption that the inner radius of the disc is at the ISCO in observation N5 is correct, {\IGR} contains a low-spin or retrograde BH, consistent with what \cite{Rao2012} suggested.

\subsection{The mass of the black-hole {\IGR}}
The inner disc radius of the accretion disc in a BH LMXBs can be measured from fitting either the disc or the reflection component. 
The inner radius of the disc, $R\rm_{dbb}$ in km, is related to the disc blackbody normalisation as:

\begin{equation}
N{\rm_{dbb}}=\frac{1}{f\rm_{col}^4}(\frac{R{\rm_{dbb}}}{D_{10}})^2 \cos i
\end{equation}
\citep{Kubota1998}, where $f\rm_{col}$ is the colour correction factor, accounting for the deviation of the local disc spectrum from a blackbody \citep*{Zhang1997}.
Assuming that the inner radii deduced from the {\sc diskbb} and {\sc relxill} components are the same, after substitution, we obtain $R\rm_{dbb}$ as:
\begin{equation}
R{\rm_{dbb}}=10^{-5}R{\rm_{in}'}\frac{GM}{c^2}=1.48R{\rm_{in}'}\frac{M}{M_{\odot}}~\rm km,
\end{equation}

where $R'\rm_{in}$ is the value of the inner radius of the disc obtained from the fits with {\sc relxill} in $R\rm_{g}$ units, $M$ is the BH mass in $\rm cgs$ units. By combining Equations 1 and 2, the mass of the BH can be rewritten as a function of the distance to the source:
\begin{equation}
\frac{M}{M_{\odot}}=A{\rm_{c}}D{\rm_{10}},  ~~~A{\rm_{c}}=\frac{f{\rm_{col}}^2}{1.48R'{\rm_{in}}}\sqrt[]{\frac{N\rm_{dbb}}{\cos{i}}}
\end{equation}
By using our best-fitting value of the inclination $i=45\degree.3\pm0\degree.7$, and $f\rm_{col}$ = 1.7 \citep{Shimura1995}, we can calculate the coefficient $A\rm_{c}$ of Equation~3 as a function of the best-fitting values of $N\rm_{dbb}$ and $R'\rm_{in}$ for each of the six observations. We show the values of $A\rm_{c}$ along the outburst in the top panel of Fig.~11; all the coefficients are consistent between different observations within errors. We therefore used the average of $A\rm_{c}$ in Equation~3 to plot the mass of the BH in {\IGR} as a function of the distance to the source in the lower panel of Fig.~11.
The gray area represents the corresponding errors at 1-$\sigma$ confidence level. 

\begin{figure} 
\centering
\resizebox{1\columnwidth}{!}{\rotatebox{0}{\includegraphics[clip]{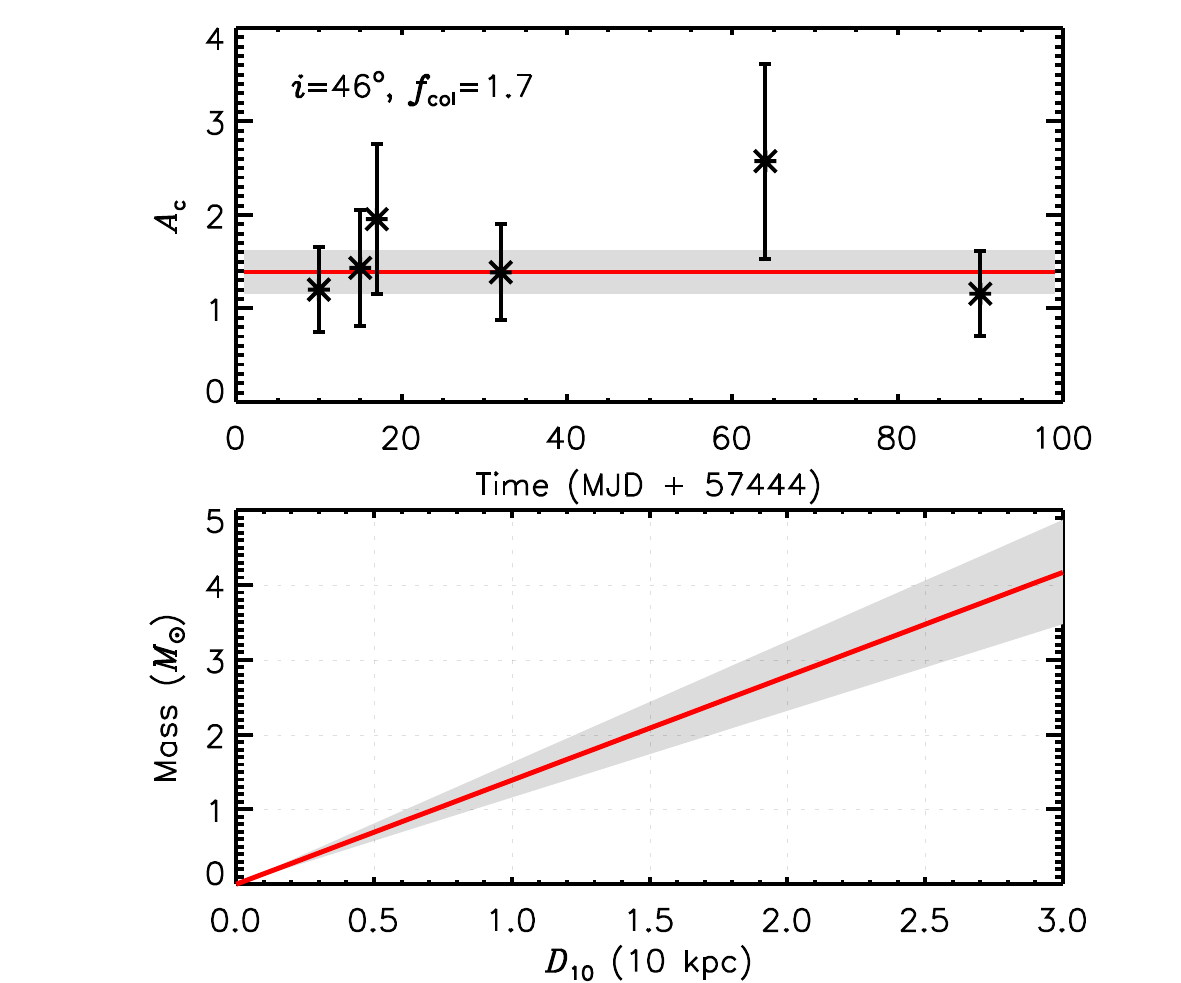}}}
\caption{Mass of {\IGR} as a function of its distance.
The six asterisks with errors correspond to the values of the coefficients in Equation~3 using the best-fitting parameters for each average observation. The red line corresponds to the average coefficient, and the gray area represents the corresponding errors with $1~\sigma$ confidence.}
\end{figure}

Based on the average mass-distance relation shown in the bottom panel of Fig.~11, 
if we assume that {\IGR} contains a $3~M_{\odot}$ black hole \citep{Altamirano2011}, the corresponding distance to {\IGR} is about 18.5$-$25.9~kpc.
For a larger BH mass, the distance to the source increases even further. This result supports the suggestion of \cite{Altamirano2011} that the BH in {\IGR}, either has a low mass, or it is very far away from us.

\subsection{Comparison between {\IGR} and {\rgs} in the heartbeat state}
Heartbeat oscillations attributed to radiation pressure instability in the accretion disc have been reported both in GRS~1915+105 (\citealt*{Greiner1996}; \citealt{Belloni1997}) and {\IGR} \citep{Altamirano2011,Janiuk2015}. 
Due to the existence of a reflection spectrum in the heartbeat state of {\IGR}, we can study the accretion geometry in this system and compare it to the case of {\rgs}.
Using simultaneous {\Nu} and {\it Chandra} observations, \cite{Zoghbi2016} carried out phase-resolved spectroscopy of the oscillations in the heartbeat state of GRS~1915+105. They extracted the {\Nu} spectra at six phases of the heartbeat QPO and fitted the spectra with the model {\sc tbabs*(ezdiskbb + relxill + cutoffpl + Gaussian)}, which is very similar to the model we used here for {\IGR}. Because {\IGR} is much weaker than {\rgs}, the data we used for phase-resolved spectroscopy here cannot constrain the reflection parameters with the same accuracy as in the case of {\rgs} presented by \cite{Zoghbi2016}.

A reflection spectrum is produced by the power-law component irradiating the surface of the accretion disc, where the X-ray photons interact with the material producing diverse atomic features (e.g., \citealt{George1991}; \citealt*{Matt1991}). If the power-law component is located close to the black hole, due to strong gravitational light bending, more of the power-law radiation would be bent towards the disc, where the reflection component originates, and less radiation would be emitted directly to the observer (e.g., \citealt*{Martocchia2000}). 

The upper panel in Fig.~12 shows the relation between the reflection flux, $F\rm_{rel}$, and the power-law flux, $F\rm_{pl}$, of the average spectra of {\IGR} in the energy range of 20$-$40~keV; in this energy range the reflection spectrum is dominated by Compton scattering, suffering weakly from the Fe abundance or ionization state of the disc \citep{Dauser2016}. 
This relation can be well fitted with a linear function which, together with the low reflection fraction, relf\_frac~=~0.21$-$0.29, indicates that the power-law emission is not strongly affected by light bending.
However, from the phase-resolved spectral analysis of {\IGR}, we found that the oscillations of the power-law flux (Fig.~10i) and the reflection flux (Fig.~10j) are uncorrelated in the energy of 3$-$75~keV; the variability of $F\rm_{pl}$ is significant whereas that of $F\rm_{rel}$ is negligible.
In the lower panel of Fig.~12, we show $F\rm_{rel}$ as a function of $F\rm_{pl}$ in the energy range of 20$-$40~keV and fitted this relation with a linear and a constant function. The fits show that the reflection flux is consistent with being constant, and the slopes of the $F\rm_{rel}$-$F\rm_{pl}$ relations for the average and the phase-resolved spectra are different at a 3.3-$\sigma$ level. 
Strangely, while the low reflection fraction values that we obtained from the fits to the phase-resolved spectra, relf\_frac~=~0.17$-$0.28, suggest that also in this case the light-bending effect is not important, the lack of correlation between the reflection and power-law fluxes appears to imply the opposite. Alternatively, the direct emission seen by the observer at infinity may not be the same as that seen by the disc during the oscillations. We tried to let the photon index vary between the reflection and the power-law components and we indeed got different values of the photon index for these two components, however, with very large uncertainties.
Further exploration of this idea is limited by the quality of our data.

\begin{figure} 
\centering
\resizebox{1\columnwidth}{!}{\rotatebox{0}{\includegraphics[clip]{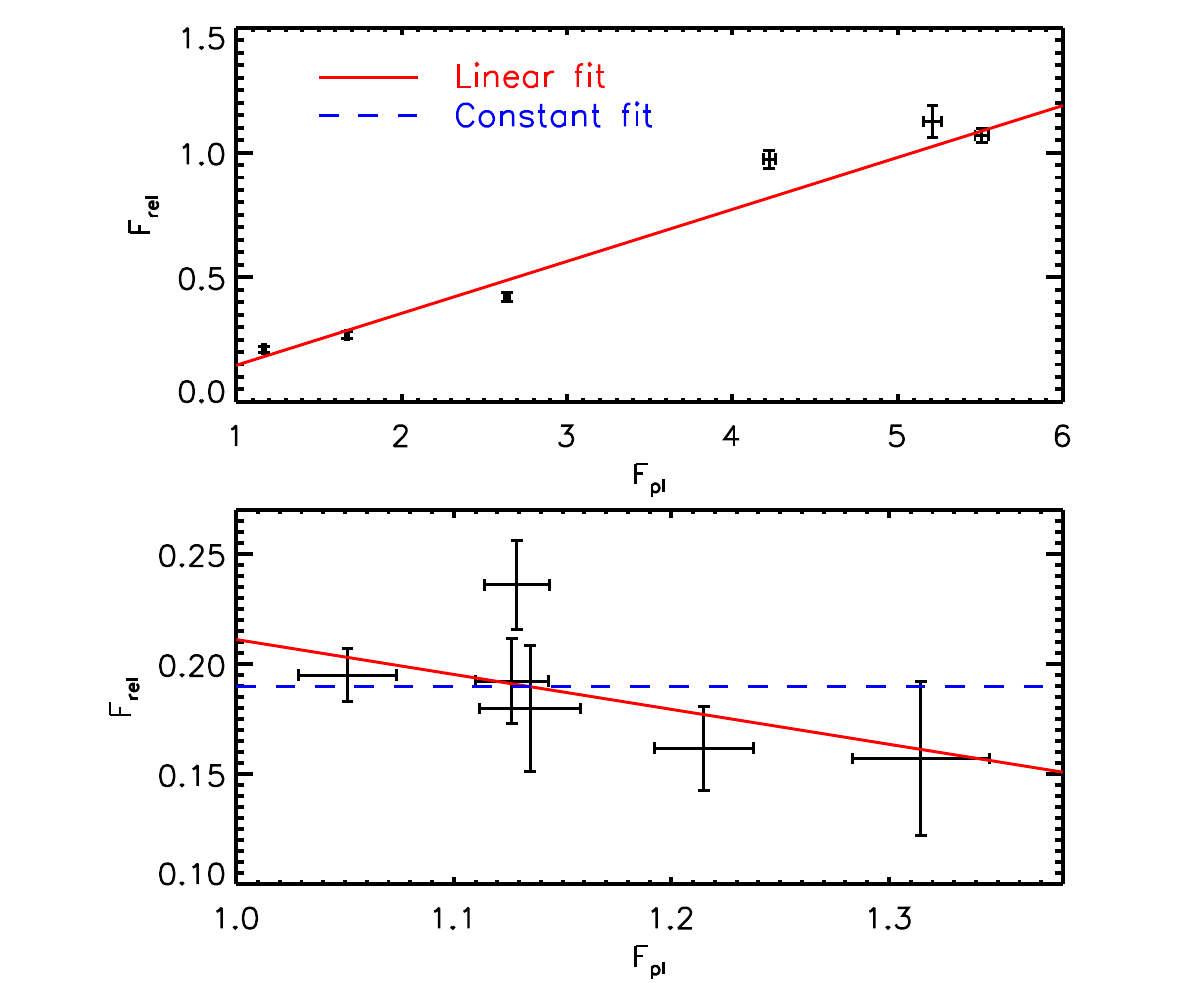}}}
\caption{The unabsorbed flux of the reflection component versus the unabsorbed flux of the power-law component for the average (upper panel) and phase-resolved (lower panel) spectra of {\IGR} in the range of 20$-$40~keV. 
All the fluxes are in units of $\rm 10^{-10} erg~cm^{-2}s^{-1}$. The solid and dashed lines show the fits of, respectively, a linear and a constant function to the data.}
\end{figure}

\cite{Zoghbi2016} found that the reflection and the Comptonised components of {\rgs} are anti-correlated. They suggested that the decrease in the power-law flux and the simultaneous increase of the reflection fraction is a result of changes of the coronal geometry. When the corona becomes small and moves close to the black hole, more photons hit the disc than reach the observer at infinity compared to the case in which the corona is high and farther away from the black hole. However, the low reflection fraction of the phase-resolved spectra of {\IGR} does not support this idea.
The steep emissivity profile, $q\rm_{in}=9.7\pm0.3$ and high inclination angle, $i=67\degree.2\pm0\degree.2$ of {\rgs} \citep{Zoghbi2016} indicate that the geometry of the accretion system in {\rgs} and in {\IGR} is different.

\section{conclusions}
The black-hole transient {\IGR} was in outburst in March 2016. From the simultaneous fits to the {\Nu} and {\sw} average spectra, 
we found that the spectral evolution of {\IGR} during this outbursts is very similar to that of 2011, except that the models that fit the spectra of the 2016 outburst require a reflection component whereas this component was not detected in the 2011 outburst.
From this reflection component,
we obtained an inclination angle of the accretion disc with respect to the line of sight, $i=45\degree.3\pm0\degree.7$, and an iron abundance $A\rm_{Fe}=3.5\pm0.3$ in units of the solar abundance, respectively. However, this high iron abundance is likely model-dependent.

We also carried out phase-resolved spectroscopy of the {\Nu} observation showing the heartbeat variability. 
The lack of correlation between the direct and reflected emission suggests that light bending is important in this case, although the low value of the best-fitting reflection fraction contradicts this scenario.
Assuming that the inner radius of the disc is located at the radius of the ISCO in {\IGR}, with 90\% confidence, the spin parameter of the BH in this source is between $-0.13$ and $0.27$.
Finally, the comparison of the reflection spectra in {\rgs} and {\IGR} indicate that the geometry of the accretion flow in these two systems is different.

\section*{Acknowledgements}
We thank our referee, Javier {Garc{\'{\i}}a}, for a
careful reading of the paper and a constructive report.
D.A acknowledges the support from the Royal Society. J.C acknowledges the support of the Science and Technology Facilities Council (STFC). A.B acknowledges the Royal Society and SERB (Science \& Engineering Research Board, India) for financial support through Newton-Bhabha Fund. This research has made use of data and/or software provided by the High Energy Astrophysics Science Archive Research Center (HEASARC), which is a service of the Astrophysics Science Division at NASA/GSFC and the High Energy Astrophysics Division of the Smithsonian Astrophysical Observatory. This research has made use of NASA's Astrophysics Data System. 

\bibliographystyle{mn2e2}
\bibliography{ref_2017}

\appendix
\section{Best-fitting parameters for the average and phase-resolved spectra of IGR~J17091--3642} 

\begin{table*} 
\caption{Best-fitting parameters of the {\Nu} and {\sw} average spectra of {\IGR} with the model {\sc phabs*(diskbb+relxill+cutoffpl)*gabs}}
 \renewcommand{\arraystretch}{1.3}
\centering
\begin{tabular}{clcccccc}
\hline \hline
\multicolumn{2}{c}{Components }& NS1 &  NS2 & NS3 & N4 & N5 & NS6 \\
{\sc phabs} &$N\rm_{H}~(10^{22}~cm^{-2})$ &\multicolumn{6}{c}{$1.50\pm0.02^{l}$}\\
\hline
{\sc diskbb}&$T\rm_{dbb}$~(keV)& $0.71\pm0.02$ &$0.70\pm0.04$ &$0.61\pm0.03$ &$0.84\pm0.01$           &$1.03\pm0.01$ &$0.84\pm0.02$ \\
&$N\rm_{dbb}$~$(R_{\rm dbb}^2/D_{10}^2 \cos~i)$&$33.1_{-3.9}^{+4.4}$ &$27.2_{-5.2}^{+6.4}$ &$66.5_{-13.2}^{+16.6}$ &$57.7\pm5.0$ &$35.9\pm1.4$ &$32.1\pm3.3$ \\
\hline                              
{\sc relxill}& $q\rm_{in}$ &\multicolumn{6}{c}{$3.7\pm0.3^{l}$}\\
& $A\rm_{Fe}$ &\multicolumn{6}{c}{$3.5\pm0.3^{l}$}\\
& $i$~($\degree$) &\multicolumn{6}{c}{$45.7\pm0.3^{l}$}\\
& $a_{*}$ &\multicolumn{6}{c}{$0.998^{f}$}\\
&$\Gamma$&$1.40\pm0.02$ &$1.47\pm0.02$ &$1.50\pm0.02$ &$2.01\pm0.02$ &$2.41\pm0.02$ &$1.95\pm0.02$ \\
&$E\rm_{cut}$~(keV) &$104.2_{-8.6}^{+7.4}$ &$86.5_{-5.7}^{+6.6}$ &$82.0_{-4.5}^{+5.1}$ &$81.6_{-4.8}^{+6.3}$ &$1000^p$ &$97.7_{-8.3}^{+10.9}$ \\
&Refl\_frac&$0.28\pm0.02$ &$0.29\pm0.03$ &$0.27\pm0.02$ &$0.21\pm0.02$ &$0.24\pm0.02$ &$0.22\pm0.03$ \\
&$R\rm_{in}$~($R\rm_{g})$&$11.2\pm1.4$ &$8.5_{-2.2}^{+1.6}$ &$9.7\pm1.7$ &$12.8_{-1.6}^{+0.9}$ &$5.4\pm1.1$ &$11.4\pm1.9$ \\
&$log~\xi$~($\rm erg~cm~s^{-1}$)&$3.04\pm0.02$ &$2.79\pm0.06$ &$2.77\pm0.05$ &$3.41\pm0.09$ &$3.52_{-0.12}^{+0.23}$ &$3.33\pm0.06$ \\
&$N\rm_{rel}~(10^{-3})$&$3.2\pm0.2$ &$3.8\pm0.4$ &$4.1\pm0.4$ &$4.2\pm0.5$ &$5.4_{-0.7}^{+1.1}$ &$2.3\pm0.2$\\ 
\hline
{\sc cutoffpl}&$N\rm_{pl}$~($\rm photons~cm^{-2}s^{-1}keV^{-1}$ at 1~keV) & $0.07\pm0.001$ &$0.11\pm0.001$ &$0.13\pm0.001$ &$0.35\pm0.02$ &$0.44\pm0.02$ &$0.17\pm0.01$ \\
\hline
{\sc gabs}&$E\rm_{gabs}$~(keV)& \multicolumn{6}{c}{$7.14\pm0.04^{l}$}\\
&$\sigma\rm_{gabs}$~(keV)& \multicolumn{6}{c}{$0.24\pm0.05^{l}$}\\
&strength ($10^{-2}$)&$6.4\pm0.6$ &$5.3\pm0.8$ &$5.6\pm0.7$ &$4.7\pm0.7$ &$4.3\pm0.9$ &$4.6\pm0.9$\\
&$\tau$ &$0.11\pm0.02$ &$0.09\pm0.02$ &$0.09\pm0.02$ &$0.08\pm0.02$ &$0.07\pm0.02$ &$0.08\pm0.02$\\
\hline
\multicolumn{2}{c}{$\chi^{2}$/d.o.f.} &\multicolumn{6}{c}{8070.6/7947}\\
\hline
\multicolumn{2}{c}{$F\rm_{dbb}~(10^{-10}~erg~cm^{-2}s^{-1})$}&$0.5\pm0.1$ &$0.4\pm0.2$ &$0.4\pm0.2$ &$2.3\pm0.3$ &$4.1\pm0.3$ &$1.2\pm0.2$ \\
\multicolumn{2}{c}{$F\rm_{rel}~(10^{-10}~erg~cm^{-2}s^{-1})$}&$0.6\pm0.1$ &$0.6\pm0.1$ &$0.6\pm0.1$ &$1.6\pm0.2$ &$1.1\pm0.2$ &$0.8\pm0.2$ \\
\multicolumn{2}{c}{$F\rm_{pl}~(10^{-10}~erg~cm^{-2}s^{-1})$}&$4.2\pm0.2$ &$6.1\pm0.3$ &$7.0\pm0.3$ &$8.3\pm0.4$ &$6.2\pm0.3$ &$4.5\pm0.3$ \\
\multicolumn{2}{c}{$F\rm_{tt}~(10^{-10}~erg~cm^{-2}s^{-1})$}&$5.4\pm0.2$ &$7.1\pm0.4$ &$8.0\pm0.4$ &$12.4\pm0.6$ &$11.3\pm0.5$ &$6.6\pm0.4$ \\
\hline
\end{tabular}
\begin{flushleft}
{\bf Note:} The symbol $l$ indicates that the parameters are linked to vary across the observations; $f$ means that the parameter is fixed during the fit; $p$ shows the parameter pegged at the upper limit. Errors are quoted at 1-$\sigma$ confidence level. All the fluxes are calculated in the energy range of 2$-$10~keV. 
\end{flushleft}
\end{table*}

\begin{table*} 
\caption{Best-fitting parameters of the phase-resolved spectra of observation N5 of {\IGR} with the model {\sc phabs*(diskbb+relxill+cutoffpl)*gabs}}

 \renewcommand{\arraystretch}{1.3}
\centering
\begin{tabular}{clcccccc}
\hline \hline
 
\multicolumn{2}{c}{Components}& ph1 &  ph2 & ph3 & ph4 & ph5 & ph6 \\
{\sc phabs} &$N\rm_{H}~(10^{22}~cm^{-2})$ &\multicolumn{6}{c}{$1.50\pm0.02^{f}$}\\
\hline
{\sc diskbb}&$T\rm_{dbb}$~(keV)& $1.11\pm0.01$ &$0.99\pm0.02$ &$1.01\pm0.02$ &$1.00\pm0.02$ &$0.99\pm0.02$ &$1.10\pm0.02$ \\
&$N\rm_{dbb}$~~$(R_{\rm dbb}^2/D_{10}^2 \cos~i)$& $30.6\pm1.5$ &$40.3\pm3.2$ &$38.3\pm2.9$ &$36.5\pm2.8$ &$39.6\pm3.2$ & $30.9\pm1.8$ \\
\hline
{\sc relxill}& $q\rm_{in}$ &\multicolumn{6}{c}{$3.68^{f}$}\\
& $A\rm_{Fe}$ &\multicolumn{6}{c}{$3.52^{f}$}\\
& $i$~($\degree$) &\multicolumn{6}{c}{$45.7^{f}$}\\
& $a_{*}$ &\multicolumn{6}{c}{$0.998^{f}$}\\
&$\Gamma$& $2.46\pm0.02$ &$2.44\pm0.02$ &$2.37\pm0.02$ &$2.38\pm0.02$ &$2.42\pm0.02$ & $2.44\pm0.02$ \\
&$E\rm_{cut}$~(keV) &\multicolumn{6}{c}{$1000^{f}$}\\
&Refl\_frac& $0.17\pm0.04$ &$0.24\pm0.02$ &$0.21\pm0.04$ &$0.23\pm0.04$ &$0.28\pm0.04$ &$0.15\pm0.04$ \\
&$R\rm_{in}$~($R\rm_{g})$& \multicolumn{6}{c}{$5.42^f$}\\
&$log~\xi$~($\rm erg~cm~s^{-1}$)& \multicolumn{6}{c}{$3.52^f$}  \\
&$N\rm_{rel}~(10^{-3})$&$6.5\pm1.4$ &$5.3\pm0.8$ &$4.3\pm0.8$ &$4.5\pm0.7$ &$5.2\pm0.7$ &$6.6\pm1.6$ \\
\hline
{\sc cutoffpl}&$N\rm_{pl}$~($\rm photons~cm^{-2}s^{-1}keV^{-1}$ at 1~keV)&$0.53\pm0.03$ &$0.44\pm0.02$ &$0.37\pm0.02$ &$0.38\pm0.02$ &$0.44\pm0.02$ &$0.54\pm0.03$ \\
\hline
{\sc gabs}&$E\rm_{gabs}$~(keV)& \multicolumn{6}{c}{$7.15\pm0.04^{l}$}\\
&$\sigma\rm_{gabs}$~(keV)&\multicolumn{6}{c}{$0.25\pm0.05^{l}$}\\
&Strength ($10^{-2}$)& \multicolumn{6}{c}{$4.1\pm0.9^{l}$}\\
&$\tau$& \multicolumn{6}{c}{$0.07\pm0.02^{l}$}\\
\hline
\multicolumn{2}{c}{$\chi^{2}$/d.o.f.} &\multicolumn{6}{c}{3581.8/3512}\\
\hline
\multicolumn{2}{c}{$F\rm_{dbb}~(10^{-10}~erg~cm^{-2}s^{-1})$}& $3.1\pm0.2$ &$2.1\pm0.1$ &$2.2\pm0.1$ &$2.1\pm0.1$ &$2.1\pm0.1$ &$3.0\pm0.2$ \\
\multicolumn{2}{c}{$F\rm_{rel}~(10^{-10}~erg~cm^{-2}s^{-1})$}&$1.0\pm0.1$ &$1.2\pm0.1$ &$1.0\pm0.1$ &$1.1\pm0.1$ &$1.4\pm0.1$ &$0.9\pm0.1$ \\
\multicolumn{2}{c}{$F\rm_{pl}~(10^{-10}~erg~cm^{-2}s^{-1})$}&$8.4\pm0.3$ &$7.2\pm0.3$ &$7.2\pm0.3$ &$7.2\pm0.4$ &$7.5\pm0.3$ &$8.9\pm0.5$ \\
\multicolumn{2}{c}{$F\rm_{tt}~(10^{-10}~erg~cm^{-2}s^{-1})$}&$12.6\pm0.4$ &$10.5\pm0.4$ &$10.6\pm0.4$ &$10.5\pm0.4$ &$11.1\pm0.4$ &$13.0\pm0.5$ \\
\hline
\end{tabular}
\begin{flushleft}
{\bf Note:} The symbol $l$ indicates that the parameters are linked to vary across the observations; $f$ means that the parameter is fixed during the fit. Errors are quoted at 1-$\sigma$ confidence level. All the fluxes are calculated in the energy range of 3$-$75~keV.
\end{flushleft}
\end{table*}

\end{document}